\documentclass[a4paper,11pt]{article}
\pdfoutput=1 

\usepackage{jcappub} 

\usepackage[T1]{fontenc}
\usepackage{centernot}

\title{Inherently stable effective field theory for dark energy and modified gravity}

\author[a]{Lucas Lombriser,}
\author[a]{Charles Dalang,}
\author[b]{Joe Kennedy,}
\author[b]{Andy Taylor}

\affiliation[a]{D\'{e}partement de Physique Th\'{e}orique, Universit\'{e} de Gen\`{e}ve, \\ 24 quai Ernest Ansermet, 1211 Gen\`{e}ve 4, Switzerland}
\affiliation[b]{Institute for Astronomy, University of Edinburgh,\\Royal Observatory, Blackford Hill, Edinburgh, EH9 3HJ, U.K.}

\emailAdd{lucas.lombriser@unige.ch}

\newcommand{\aM}{\alpha_{\rm M}}
\newcommand{\aB}{\alpha_{\rm B}}
\newcommand{\aK}{\alpha_{\rm K}}
\newcommand{\cT}{c_{\rm T}}
\newcommand{\cs}{c_{\rm s}}
\newcommand{\aMtoday}{\alpha_{{\rm M}0}}
\newcommand{\aBtoday}{\alpha_{{\rm B}0}}
\newcommand{\tB}{\tilde{B}}
\newcommand{\tBtoday}{\tilde{B}_0}

\newcommand{\rhom}{\rho_{\rm m}}
\newcommand{\rhoDE}{\rho_{\rm DE}}
\newcommand{\rhoeff}{\rho_{\rm eff}}
\newcommand{\Om}{\Omega_{\rm m}}
\newcommand{\ODE}{\Omega_{\rm DE}}
\newcommand{\ODEtoday}{\Omega_{{\rm DE}0}}
\newcommand{\Omtoday}{\Omega_{{\rm m}0}}
\newcommand{\weff}{w_{\rm eff}}

\newcommand{\aacc}{a_{\rm acc}}
\newcommand{\chiacc}{\chi_{\rm acc}}

\abstract{
The growing wealth of cosmological observations places succeedingly more stringent constraints on dark energy and alternative gravity models. To most effectively exploit this data and infer the broadest of implications for the vast conceivable model space, the development of unified frameworks for generalised cosmological predictions has become an important endeavour.
Particularly successful for this purpose has been the effective field theory of dark energy and modified gravity. In its practical application the formalism is however still restrained by questions surrounding the adequate parametrisation of the free time-dependent functions inherent to the framework, which should respect a multitude of requirements, ranging from simplicity, generality, and representativity of known theories to the computational efficiency. But in particular, for theoretical soundness, the parameter space should adhere to strict stability requirements.
Focussing on Horndeski gravity with luminal speed of gravity, we explore the inherently stable effective field theory that we have recently introduced with the physical basis functions of Planck mass evolution, sound speed of the scalar field fluctuation, kinetic coefficient, and background expansion.
We devise a parametrisation of the basis functions that can straightforwardly be configured to evade theoretical pathologies such as ghost or gradient instabilities or to accommodate further theoretical priors such as a luminal or subluminal scalar sound speed. The parametrisation is simple yet general, conveniently represents a broad range of known dark energy and gravitational theories, and with a simple additional approximation can be rendered numerically highly efficient. Finally, by operating in our new basis, we show that there are no general limitations from stability requirements on the current values that can be assumed by the phenomenological modification of the Poisson equation and the gravitational slip besides the exclusion of anti-gravity. The inherently stable effective field theory is ready for implementation in parameter estimation analyses employing current or future cosmological observations amenable to linear perturbation theory.
}

\begin{document}
\maketitle
\flushbottom

\section{Introduction} \label{sec:intro}

Cosmological applications of General Relativity (GR) involve vastly different length scales to the Solar System, where our Theory of Gravity has been put under intense scrutiny and so far successfully passed all tests~\cite{Will2014}. In orders of magnitude this extrapolation is comparable to a projection in scale from the diameter of an atomic nucleus to distances inherent to the realm of everyday human experience. It is therefore worthwhile to perform independent tests of GR in the cosmological regime. Additional motivation for a thorough examination of cosmological gravity can be drawn from the necessity of a large dark sector in the energy budget of our Universe to explain the large-scale observations in the context of GR. In particular the observed late-time accelerated expansion of our Cosmos has traditionally been an important driver for the development of alternative theories of gravity~\cite{Koyama2016,Joyce2016,Ishak:2018his,Heisenberg:2018vsk}. While generally attributed to a cosmological constant $\Lambda$, thought to arise from vacuum fluctuations, quantum theoretical calculations deviate from the observed value by more than $50$ orders of magnitude~\cite{Weinberg:1989,Martin:2012bt} (see, however, Refs.~\cite{Kaloper:2013zca,Wang:2017oiy,Lombriser:2018aru}). It was therefore conjectured that the cosmic acceleration could instead be due to a dark energy field that permeates the Universe and comes to dominate the dynamics of its expansion at late times~\cite{Copeland:2006wr}. It is further hypothesised that perhaps this field could be linked to our lack of understanding of gravitational physics in the ultraviolet (UV), where a more fundamental theory may give rise to a remnant extra degree of freedom in the infrared that accelerates the expansion. This field could furthermore be coupled to the metric and modify the gravitational interactions. The measured luminal speed of gravity at late times~\cite{Abbott2017b}, however, challenges the concept that such a modification could be directly responsible for cosmic acceleration~\cite{Lombriser2016a}. Nevertheless, the dark energy field could still be coupled to the metric or the matter fields in a universal or nonuniversal manner that can manifest in cosmological observations. The past decades have seen a steep growth in high-quality cosmological data and succeedingly more stringent constraints on dark energy and modified gravity models have been inferred from measurements of the background evolution, the cosmic microwave background, the large-scale structure, and the propagation of gravitational waves (see Refs.~\cite{Copeland:2006wr,Lombriser2014a,Koyama2016,Joyce2016,Ishak:2018his,Kase:2018aps} for reviews).

In order to most effectively make use of this wealth of observations to infer the broadest of implications for the vast range of dark energy and modified gravity models conceivable, it has been of great interest to develop unified frameworks for making generalised predictions for this space of models (see, e.g., Ref.~\cite{Lombriser:2018} for a review). Particularly successful has been the effective field theory (EFT) formalism for dark energy and modified gravity~\cite{Gleyzes:2014rba}. However, in its practical application of the comparison of generalised predictions to observations, the formalism is still plagued by questions of the adequate parametrisation of the free functions inherent to the framework. Such a parametrisation should ideally be simple yet representative for known models, accurately recovering them for specific parameter values. It should furthermore allow a restriction to a parameter region that does not suffer from theoretical pathologies, and it should be efficient in its prediction of cosmological observables for the numerical applications. It is nontrivial to simultaneously meet all of these requirements. Particularly, questions surrounding the stability of the theory space have inspired much work~\cite{Linder:2015rcz,DeFelice:2016ucp,Linder:2016wqw,Gleyzes:2017kpi,Kreisch:2017uet,Peirone:2017ywi,Kennedy:2018gtx,Denissenya:2018mqs}. To evade the sampling of unstable models by default, in Ref.~\cite{Kennedy:2018gtx}, we have therefore proposed the adoption of a manifestly stable EFT basis.

In this paper, we explore to what extent this new EFT basis can conveniently represent a broad range of known dark energy and gravitational theories and can then furthermore be parametrised in its time dependence in a straightforward fashion to provide a simple but general representation of the available model space.
We also inspect how efficient predictions can be obtained that can then be used in parameter estimation analyses.
Hereby, we restrict to the EFT operators that are introduced by a new scalar degree of freedom that may interact with the gravitational tensor field while yielding at most second-order equations of motion, hence, is embedded in the Horndeski action~\cite{Horndeski1974}, while also adhering to a luminal speed of gravitational waves.

The paper is organised as follows.
In Sec.~\ref{sec:ISB}, we provide a brief review of a number of aspects of Horndeski scalar-tensor theories, the effective field theory of modified gravity and dark energy, and the stable basis that will be important to our analysis.
We then show in Sec.~\ref{sec:models} how a range of well-studied modified gravity and dark energy models is conveniently mapped onto the inherently stable EFT.
A parametrisation of the basis under consideration of generality, theoretical stability and representativity as well as numerical efficiency is designed in Sec.~\ref{sec:parametrising}.
We also discuss some phenomenological aspects of directly parametrising the stable basis.
Finally, we conclude with a discussion of our results in Sec.~\ref{sec:conclusions} and present a number of relations for mapping Horndeski scalar-tensor theories onto the stable EFT as well as a reconstruction of the class of Horndeski models corresponding to a particular set of basis functions in Appendix~\ref{sec:mappings}.

\section{Inherently stable effective field theory} \label{sec:ISB}

Before engaging in the construction of a theoretically healthy basis for the parametrisation and exploration of the broad space of modified gravity and dark energy models based on an extra scalar degree of freedom, we briefly review some important aspects of the Horndeski scalar-tensor theory in Sec.~\ref{sec:Horndeski} and the EFT formalism in Sec.~\ref{sec:EFT}.
This is mainly intended for the unfamiliar reader and for fixing notation.
In Sec.~\ref{sec:EFTISB}, we then discuss the new stable EFT basis introduced in Ref.~\cite{Kennedy:2018gtx}.
We furthermore refer the reader to Appendix~\ref{sec:mappings} for the relations connecting the different frameworks.

\subsection{Horndeski scalar-tensor gravity} \label{sec:Horndeski}

Horndeski gravity~\cite{Horndeski1974,Deffayet:2011gz,Kobayashi:2011nu} describes the most general, four-dimensional, Lorentz-covariant scalar-tensor theory that produces at most second-order equations of motion.
Its action is given by
\begin{equation}
 S = \int d^4x \sqrt{-g} \left[ \sum_{i=2}^5 \mathcal{L}_i + \mathcal{L}_{\rm m}(g_{\mu\nu},\psi_{\rm m}) \right] \,, \label{eq:horndeski}
\end{equation}
where the different Lagrangian densities are specified by
\begin{eqnarray}
 \mathcal{L}_2 & = & G_2(\phi,X) \,, \\
 \mathcal{L}_3 & = & G_3(\phi,X)\Box \phi \,, \\
 \mathcal{L}_4 & = & G_4(\phi,X)R + G_{4X}(\phi,X) \left[ (\Box\phi)^2 - (\nabla_{\mu}\nabla_{\nu}\phi)^2 \right] \,, \\
 \mathcal{L}_5 & = & G_5(\phi,X) G_{\mu\nu}\nabla^{\mu}\nabla^{\nu}\phi - \frac{1}{6} G_{5X}(\phi,X) \left[ (\Box\phi)^3 - 3\Box\phi(\nabla_{\mu}\nabla_{\nu}\phi)^2 + 2 (\nabla_{\mu}\nabla_{\nu}\phi)^3 \right]
\end{eqnarray}
with $X\equiv-\frac{1}{2}\partial_{\mu}\phi\partial^{\mu}\phi$ and minimally coupled matter fields $\psi_{\rm m}$ in $\mathcal{L}_{\rm m}$.
$R$ and $G_{\mu\nu}$ denote the Ricci scalar and Einstein tensor of the Jordan frame metric $g_{\mu\nu}$.
For convenience, we shall adopt units where the Planck mass $M_{\rm Pl}$ and the speed of light in vacuum are set to unity.
While the Horndeski action is constructed to evade Ostrogradsky instabilities, it does not imply that a given background solution is stable under fluctuations for any choice of the $G_i$ functions (Sec.~\ref{sec:EFT}).

In our discussion, we shall furthermore adopt a luminal speed of gravity over the entire cosmic history, motivated by the constraint on the speed of gravitational waves $|\cT^2-1|\lesssim10^{-15}$ at redshifts of $z\lesssim0.01$~\cite{Abbott2017b} by a LIGO/Virgo gravitational wave with electromagnetic counterparts emitted by a neutron star merger (also see Refs.~\cite{Nishizawa2014,Lombriser2015c}).
In particular, the constraint implies that~\cite{McManus2016}
\begin{equation}
 G_{4X} \simeq G_5 \simeq 0
\end{equation}
for the low-redshift regime (see Refs.~\cite{Creminelli:2018xsv,Kase:2018aps} for recent and more general discussions).
As a consequence, Eq.~\eqref{eq:horndeski} can no longer provide a cosmic self-acceleration by a genuine modification of gravity that would be compatible with observations of the large-scale structure~\cite{Lombriser2016a} (also see Sec.~\ref{sec:msa}).
Note, however, that there could be a scale dependence in $\cT$~\cite{Battye2018,deRham:2018red} arising from completion of the effective theory in the UV, recovering $\cT=1$ for the energy scales relevant to current direct gravitational wave measurements.
In this paper, we focus on the late-time Universe, and leaving UV completion effects aside we specify to the contributions of $G_4(\phi)$, $G_3(\phi,X)$, and $G_2(\phi,X)$ only.

\subsection{Effective field theory of dark energy and modified gravity} \label{sec:EFT}

A systematic approach to covering the broad cosmological phenomenology of the range of possible dark energy and modified gravity models embedded in the Horndeski action~(\ref{eq:horndeski}) is employed with the effective field theory of dark energy and modified gravity~\cite{Creminelli2009,Park2010,Battye2012,Gubitosi2012,Bloomfield2012,Gleyzes2013,Hu:2013twa,Tsujikawa2014,Bellini2014,Lombriser2014,Lombriser2015b,Zumalacarregui:2016pph,Lagos2017} (see Ref.~\cite{Gleyzes:2014rba} for a review).
Hereby the gravitational action is cast into unitary gauge, where the time coordinate is set to absorb the scalar field perturbation in the metric $g_{\mu\nu}$.
A generalised gravitational action can then be built from the combination of geometric quantities that are invariant under time-dependent spatial diffeomorphisms with free time-dependent coefficients.
To quadratic order, which is sufficient for describing the cosmological background evolution and linear fluctuations generated by Eq.~\eqref{eq:horndeski}, one obtains for a low-energy, four-dimensional, Lorentz-covariant effective theory with at most second-order derivatives in the equations of motion that satisfy $\cT=1$,
\begin{eqnarray}
 S_g & = & \frac{1}{2} \int d^4x\sqrt{-g}\left\{ \vphantom{R^{(3)}\hat{M}^2m_2^2} \Omega(t) R - 2\Lambda(t) - \Gamma(t) \delta g^{00} + M_2^4(t) (\delta g^{00})^2 - \bar{M}_1^3(t) \delta g^{00} \delta K^{\mu}_{\ \mu} \right\}, \label{eq:eftaction}
\end{eqnarray}
where the total action $S=S_g + S_{\rm m}(g_{\mu\nu},\psi_{\rm m})$ and we have adopted the notation of Ref.~\cite{Lombriser2014}.
$K_{\mu\nu}$ denotes the extrinsic curvature tensor and $\delta$ indicates perturbations around the background.
$\Lambda$CDM is recovered for $\Omega=1$, constant $\Lambda$, and all remaining coefficients vanishing.
For quintessence models~\cite{Ratra:1987rm,Wetterich1988}, $\Omega=1$ with $\Lambda$ and $\Gamma$ describing the scalar field potential and kinetic terms.
$M_2^4\neq0$ is introduced in k-essence~\cite{ArmendarizPicon1999} and $\bar{M}_1^3\neq0$ in the cubic Galileon~\cite{Nicolis2008} and Kinetic Gravity Braiding (KGB)~\cite{Deffayet2010} models.
All the coefficients are used to embed Horndeski theories with $\cT=1$ (Sec.~\ref{sec:Horndeski}).

There is a total of five coefficients in the action~\eqref{eq:eftaction}, where the scale factor $a(t)$ or the Hubble function $H(t)$ of the spatially homogeneous and isotropic background adds a sixth function.
For simplicity spatial flatness and a matter-only universe with pressureless dust of energy density $\rhom$ is assumed throughout the paper. The Friedmann equations
\begin{eqnarray}
  H^2\left(1 + \frac{\Omega'}{\Omega}\right) & = & \frac{\kappa^2\rhom + \Lambda + \Gamma}{3\Omega} \,, \label{eq:friedmann1} \\
  \left(H^2\right)' \left(1 + \frac{1}{2}\frac{\Omega'}{\Omega}\right) + H^2\left(3 + \frac{\Omega''}{\Omega} + 2\frac{\Omega'}{\Omega}\right) & = & \frac{\Lambda}{\Omega} \,, \label{eq:friedmann2}
\end{eqnarray}
follow from variation of the action with respect to the metric and provide two constraints between the first three background coefficients in Eq.~\eqref{eq:eftaction}.
Primes denote derivatives with respect to $\ln a$ here and throughout the paper.
Hence, for specified matter content and spatial curvature, the space of dark energy and modified gravity models spanned by Eq.~\eqref{eq:eftaction}, and therefore the cosmological background evolution and perturbations arising from the action~\eqref{eq:horndeski}, is characterised by four free functions of time.

Similarly to Eq.~\eqref{eq:eftaction}, the effective action can also be built from the geometric quantities introduced in an Arnowitt-Deser-Misner (ADM)~\cite{Arnowitt:1959ah} 3+1 decomposition of spacetime, where the uniform scalar field hypersurfaces correspond to constant time hypersurfaces~\cite{Gleyzes2013,Bellini2014}.
Variation of the action then defines an equivalent set of functions for describing the cosmological phenomenology to linear order.
The formalism separates out the expansion history $H$ as the free function for the background, which relates to the EFT functions $\Omega$, $\Gamma$, and $\Lambda$ through the Friedmann equations~(\ref{eq:friedmann1}) and (\ref{eq:friedmann2}).
The three time-dependent functions that then characterise the linear cosmological fluctuations are~\cite{Bellini2014,Gleyzes:2014rba}:
\begin{eqnarray}
 \aK & \equiv & \frac{\Gamma+4M_2^4}{H^2M^2} \,, \label{eq:aK} \\
 \aB & \equiv & \frac{H\Omega' + \bar{M}_1^3}{2 H M^2} \,, \label{eq:aB} \\
 \aM & \equiv & \frac{(M^2)'}{M^2} \,. \label{eq:aM}
\end{eqnarray}
The kineticity $\aK$ parametrises the contribution of a kinetic energy of the scalar field, the braiding function $\aB$ describes its kinetic braiding or mixing with the metric field, where we adopt the notation of Ref.~\cite{Gleyzes:2014rba}, and $\aM$ denotes the evolution rate of the gravitational coupling or the effective Planck mass $M^2 \equiv \Omega$.
$\Lambda$CDM is recovered when $\alpha_i=0$ $\forall i$ and we will discuss further examples of gravitational models in Sec.~\ref{sec:models}.

The perturbed modified Einstein and scalar field equations as well as a reduced system of differential equations in this formalism can be found in Refs.~\cite{Gleyzes:2014rba,Bellini2014,Lombriser2015b}.
In particular, the modified Einstein equations can be combined such that effectively only two time and scale dependent functions characterise the modifications of the large-scale structure, on top of the modified background expansion.
These are an effective modification of the Poisson equation $\mu(a,k)$ and a gravitational slip, or effective anisotropic stress, $\gamma(a,k)$, where $k$ denotes the Fourier space wavenumber.
More specifically,
\begin{eqnarray}
 k_H^2\Psi & = & -\frac{\kappa^2\rhom}{2H^2}\mu(a,k)\Delta_{\rm m} \,, \label{eq:mu} \\
 \Phi & = & -\gamma(a,k)\Psi \,, \label{eq:gamma}
\end{eqnarray}
where the amplitudes of the scalar fluctuations around the Friedmann-Lema\^itre-Robertson-Walker (FLRW) background metric are cast in Newtonian gauge with $\Psi\equiv\delta g_{00}/(2g_{00})$ and $\Phi\equiv\delta g_{ii}/(2g_{ii})$, the amplitude of the matter density perturbation $\Delta_{\rm m}$ is specified in the comoving gauge, and $k_H\equiv k/(aH)$.
$\Lambda$CDM is recovered for $\mu=\gamma=1$.
For Horndeski theories, at leading order in $k$, formally corresponding to the linear limit of $k\rightarrow\infty$, time derivatives of the metric potentials and large-scale velocity flows do not appear~\cite{Lombriser2015b} and hence can be neglected with respect to spatial derivatives and matter density fluctuations at sufficiently small linear scales.
One then obtains for the effective modifications
\begin{eqnarray}
 \mu_{\infty} & = & \frac{1}{M^2}\frac{2 (\aB-\aM)^2 + \alpha c_{\rm s}^2}{\alpha \cs^2} \,, \label{eq:muQS} \\
 \gamma_{\infty} & = & \frac{2\aB (\aB-\aM) + \alpha c_{\rm s}^2}{2 (\aB-\aM)^2 + \alpha \cs^2} \,, \label{eq:gammaQS}
\end{eqnarray}
where $\alpha \equiv 6 \aB^2 + \aK$ and the sound speed of the scalar mode is specified by the relation
\begin{equation}
 \cs^2 = -\frac{2}{\alpha} \left[ \aB' + (1 + \aB)^2 - \left(1 + \aM - \frac{H'}{H}\right)(1 + \aB) + \frac{\rhom}{2H^2 M^2} \right] \,. \label{eq:cs}
\end{equation}
It is worth noting that the combination of $\alpha\cs^2$ cancels the contribution of $\alpha$ or $\aK$, and hence $M_2^4$, in Eqs.~\eqref{eq:muQS}--\eqref{eq:gammaQS}.
However, $\aK$ can give rise to a clustering effect on near-horizon scales, which may for instance manifest in a multi-tracer analysis of galaxy redshift survey data~\cite{Lombriser2013a}.
Whereas $\aB$, $\aM$, and $H$ are tested by the subhorizon large-scale structure and background data with no fundamental degeneracies~\cite{Lombriser2014,Lombriser2015c}, the ultra-large scale tests are crucial to constrain the remaining EFT freedom $\aK$ that is otherwise not observable.
For this purpose, one can extend Eqs.~\eqref{eq:muQS}--\eqref{eq:gammaQS} in a semidynamical expansion~\cite{Lombriser2015b} to lower orders in $k$.
Finally, the EFT function $\aK$ or $\alpha$ is also relevant for formulating stable modified gravity and dark energy theories (Sec.~\ref{sec:EFTISB}).

Besides affecting the clustering and lensing of the large-scale structure, an evolving Planck mass with the rate $\aM$, also affects the cosmological propagation of gravitational waves, where (see, e.g.,~\cite{Gleyzes:2014rba,Saltas2014,Lombriser2015c,Nishizawa2017})
\begin{equation}
 h_{ij}'' + \left(3 + \aM + \frac{H'}{H} \right)h_{ij}' + k_H^2 h_{ij} = 0 \label{eq:gw}
\end{equation}
with $h_{ij}\equiv g_{ij}/g_{ii}$ denoting the linear traceless spatial tensor perturbation.
It was pointed out in Ref.~\cite{Saltas2014} that for low-redshift modifications the additional damping that the wave experiences on top of the Hubble friction can be tested with Standard Sirens~\cite{Schutz1986,Holz2005,Abbott2017c}, and a first constraint of $|M^2(z=0)-1|\lesssim3.5\times10^{-3}$ was estimated in Ref.~\cite{Lombriser2015c} from forecasts of Standard Sirens detected by the Laser Interferometer Space Antenna (LISA)~\cite{Amaro-Seoane:2017}, and comparable constraints have been inferred for the Einstein Telescope (ET) in Ref.~\cite{Belgacem:2018lbp}.

The EFT description of the cosmological background and linear perturbations discussed here can furthermore be connected to a generalised phenomenological parametrisation of the screening mechanisms encountered in modified gravity theories~\cite{Lombriser2016b} or post-Newtonian and post-Einsteinian expansions describing these modifications~\cite{McManus2017,Nishizawa2017,Bolis:2018kcq} (see Ref.~\cite{Lombriser:2018} for a review).

Finally, for the cosmological background of the modified gravity and dark energy models described by the EFT formalism to be stable towards scalar and tensor fluctuations, the $\alpha_i$ functions must satisfy~\cite{Bellini2014}
\begin{equation}
 \frac{M^2\alpha}{(1+\alpha_{\rm B})^2} > 0 \,, \quad c_{\rm s}^2>0 \,, \quad M^2>0 \,, \label{eq:stability}
\end{equation}
which also implies $\alpha>0$.

\subsection{Stable effective field theory basis} \label{sec:EFTISB}

In order to explore the broad phenomenology available to EFT modifications, a range of different time parametrisations have been adopted for the $\alpha_i$ functions in Eqs.~\eqref{eq:aK}--\eqref{eq:aM}, or the EFT coefficients in action~\eqref{eq:eftaction}, with the ultimate aim of performing as generic a test of gravity and dark energy as possible (see, e.g., discussions in Refs.~\cite{Hu:2013twa,Linder:2015rcz,DeFelice:2016ucp,Linder:2016wqw,Zumalacarregui:2016pph,Gleyzes:2017kpi,Kreisch:2017uet,Peirone:2017ywi,Kennedy:2018gtx,Denissenya:2018mqs}).
Importantly, these parametrisations have typically not been devised to a priori satisfy the stability criteria in Eq.~(\ref{eq:stability}).
As a consequence the sampling of theoretically healthy parameter regions, for example when conducting a Markov Chain Monte Carlo (MCMC) analysis to constrain the available EFT space with observations~\cite{Hu:2013twa,Zumalacarregui:2016pph}, can be highly inefficient with only a small fraction of the samples hitting a stable region of parameter space (see, e.g., Ref.~\cite{Perenon:2015sla}).
This can produce complicated contours on the viable parameter space that are statistically difficult to interpret.
For instance, it may leave $\Lambda$CDM confined to a narrow corner of two intersecting edges produced by the stability requirements that may yield spurious evidence against standard cosmology due to the sparse sampling.

To circumvent such issues, in Ref.~\cite{Kennedy:2018gtx} we have proposed that the functions to be sampled should be the three stability conditions~\eqref{eq:stability} themselves instead of the three $\alpha_i$ functions in Eqs.~\eqref{eq:aK}--\eqref{eq:aM} or a choice of three independent EFT coefficients in action~\eqref{eq:eftaction}.
More specifically, a manifestly stable basis for the space of EFT modifications spanned by Eq.~\eqref{eq:eftaction} that describes the freedom in the cosmological background and linear perturbations of the theoretically healthy subset of models embedded in the Horndeski action~\eqref{eq:horndeski}, is provided by the stability functions~\cite{Kennedy:2018gtx}
\begin{equation}
 M^2 > 0 \,, \quad \cs^2 > 0 \,, \quad \alpha > 0 \,, \label{eq:basis}
\end{equation}
along with a boundary condition $\aBtoday = const.$ and the background expansion history $H$.
The boundary condition is required to connect the basis~\eqref{eq:basis} to the $\alpha_i$ functions in Eqs.~\eqref{eq:aK}--\eqref{eq:aM} and the effective modifications in Eqs.~\eqref{eq:muQS}--\eqref{eq:gammaQS}, which requires the integration of Eq.~\eqref{eq:cs} to derive the braiding function $\aB$, where $\aK$ is subsequently obtained from $\aB$ and $\alpha$.

The validity of adopting this new basis is best examined by defining a function $\tB$ through
\begin{equation}
1+\aB\equiv\frac{\tB'}{\tB} \,.
\end{equation}
From the relation of sound speed to the $\alpha_i$ functions in Eq.~\eqref{eq:cs}, it then follows that $\tB$ satisfies the linear homogeneous second-order differential equation
\begin{equation}
 \tB'' - \left(1+\aM-\frac{H'}{H} \right) \tB' + \left(\frac{\rhom}{2H^2 M^2} + \frac{\alpha \cs^2}{2} \right) \tB = 0 \,. \label{eq:secondorderODE}
\end{equation}
By the existence and uniqueness theorem for ordinary differential equations, Eq.~\eqref{eq:secondorderODE} is guaranteed to have a real solution given real boundary conditions on $\tB$ and $\tB'$, for instance, specified by $\aBtoday$ with arbitrary normalisation of the amplitude of $\tB$.
The stability functions can therefore be mapped uniquely onto the $\alpha_i$ functions.

In Ref.~\cite{Kennedy:2018gtx}, we advocated the adoption of the stable basis for inferring observational constraints on the space of EFT modifications to avoid the range of issues introduced by stability checks of sampling $\alpha_i$ parametrisations.
It was correctly remarked in Ref.~\cite{Denissenya:2018mqs} that the need for solving Eq.~\eqref{eq:secondorderODE} compensates for the speed-up by the stable sampling.
As we shall discuss in Sec.~\ref{sec:approximations}, this problem can be circumvented with a suitable approximation for the solutions of $M^2$ or $\aB$ for a given parametrisation of the basis~\eqref{eq:basis}.
A further advantage of adopting the stable basis is that it provides an immediate physical interpretation of the observational constraints.
Parametrisations in $H$ classify quintessence dark energy models, for which $\alpha>0$.
An additional departure from $\cs^2=1$ can then be attributed to a more exotic dark energy, where $\aBtoday\neq0$ adds an imperfection to the dark fluid.
Finally, an evolution of $M^2$ indicates a genuine modification of gravity.
In contrast, $\Lambda$CDM is recovered for $M^2=1$, $\alpha=\aBtoday=0$ with arbitrary $\cs^2$ that can be marginalised over.
Furthermore, $H^2 = H_{\Lambda{\rm CDM}}^2 \equiv (\rhom + \Lambda)/3$ with cosmological constant $\Lambda$.
Finally, it is also worth noting that the new basis addresses the measure problem on the parameter space with clearer physical motivation for prior ranges on the parameters than for $\alpha_{i}$ and allowing bounds to be placed on the desired accuracy of the measurements for a particular set of parameter values.

\section{Model landscape} \label{sec:models}

Having briefly reviewed Horndeski scalar-tensor theory, the EFT of dark energy and modified gravity and the stable basis in Sec.~\ref{sec:ISB}, we shall now discuss how a range of frequently studied submodels of Horndeski gravity straightforwardly map onto our new EFT basis.
The representation of $\Lambda$CDM was specified in Sec.~\ref{sec:EFTISB},
and in the following we discuss the examples of quintessence and k-essence models (Sec.~\ref{sec:darkenergy}), minimally coupled cubic Galileon and KGB models (Sec.~\ref{sec:KGB}), Brans-Dicke scalar-tensor theories supplied with arbitrary potential and kinetic terms (Sec.~\ref{sec:JBD}), the minimal self-acceleration model~\cite{Lombriser2016a,Lombriser:2018} and no-slip gravity~\cite{Linder:2018jil} (Sec.~\ref{sec:msa}) as well as Horndeski theories with luminal speed of gravitational waves (Sec.~\ref{sec:luminalhorndeski}).
A summary of the representation of this model landscape is given in Table~\ref{tab:modellandscape}.

\begin{table}
    \centering
    \begin{tabular}{l|c|c|c|c|c|c}
      & $\Lambda$CDM & quintessence & k-essence & $f(R)$ & JBD & min/no-$\gamma$ \\
     \hline
     $M^2$ & $1$ & $1$ & $1$ & $1+f_R$ & $\phi$ & $M^2$ \\
     $\alpha$ & $0$ & $3\ODE(a)[1+w(a)]$ & $\frac{3\ODE(a)[1+w(a)]}{\cs^2}$ & $\frac{3}{2}\left(\frac{f_R'}{1+f_R}\right)^2$ & $\frac{2\omega+3}{2}\left(\frac{\phi'}{\phi}\right)^2$ & $\alpha$ \\
     $\cs^2$ & -- & $1$ & $\cs^2$ & $1$ & $1$ & $\cs^2$ \\
     $\aBtoday$ & $0$ & $0$ & $0$ & $\left.\frac{1}{2}\frac{f_R'}{1+f_R}\right|_0$ & $\left.\frac{1}{2}\frac{\phi'}{\phi}\right|_0$ & $\left.\frac{(M^2)'}{M^2}\right|_0$
    \end{tabular}
    \caption{Mapping of $\Lambda$CDM and a few frequently studied dark energy and modified gravity models onto the stable effective field theory basis~\eqref{eq:basis}.
    Excluded here is the function $H$, which for $\Lambda$CDM and the minimal self-acceleration model (min) is given by $H_{\Lambda{\rm CDM}}$ and otherwise is determined by the given model functions, or replaces one of the given model functions.
    Only in full $\cT=1$ Horndeski gravity, it becomes an independent free function.
    No-slip gravity (no-$\gamma$) generalises the minimal self-acceleration model to a free evolution of $M^2$, where $G_3=\ln\sqrt{X}$ in Brans-Dicke representation.
    For the minimal model, $M^2$ is instead directly specified by the expansion history.
    Also see Sec.~\ref{sec:KGB} for the expressions for the minimally coupled cubic Galileon model with $X\Box\phi$ term and Kinetic Gravity Braiding.
    Note that $f(R)$ gravity is reproduced in the limit of Jordan-Brans-Dicke (JBD) gravity with Brans-Dicke parameter $\omega(\phi)=0$.
    }
    \label{tab:modellandscape}
\end{table}

\subsection{Quintessence and k-essence} \label{sec:darkenergy}

Quintessence models~\cite{Ratra:1987rm,Wetterich1988} are the archetypal dark energy models, where a scalar field is introduced with a potential that drives the late-time accelerated expansion of our Universe.
More specifically, $G_2(\phi,X)=X+V(\phi)$, $G_3=0$, and $G_4=1/2$ in the action~\eqref{eq:horndeski}.
The background expansion history for quintessence models can be written as
\begin{equation}
 H^2 = \frac{\rhom}{3} + \frac{\rhoDE}{3} \,,
\end{equation}
where the dark energy density $\rhoDE$ satisfies the energy conservation equation
\begin{equation}
 \rhoDE' = -3[1+w(a)]\rhoDE \,,
\end{equation}
with dark energy equation of state $w(a)$.
Due to $G_4=1/2$ and $G_3=0$, we have $M^2=1$ and $\aB=0$.
The kinetic contribution in $G_2$ then gives rise to~\cite{Lombriser2015b}
\begin{equation}
 \alpha = \aK = 3[1-\Om(a)][1+w(a)] = 3\ODE(a)[1+w(a)] \,,
\end{equation}
where $\Om(a)\equiv H_0^2\Omtoday a^{-3}/H^2$, $\ODE(a)\equiv 1-\Om(a)$, and $H_0 \equiv H(z=0)$ denotes the Hubble constant.
It furthermore follows from Eq.~(\ref{eq:cs}) that
\begin{equation}
 \cs^2 = 1 \,,
\end{equation}
using that
$-(2/3)(H'/H) = \Om(a) + [1+w(a)][1-\Om(a)]$.
We observe that due to the no-ghost condition $\alpha>0$, one must have
\begin{equation}
 w(a) > -1 \,,
\end{equation}
which recovers the well-known result that quintessence models cannot give rise to a phantom dark energy equation of state with $w<-1$.

We can generalise the $G_2$ Horndeski function to include noncanonical kinetic contributions or even leave $G_2(\phi,X)$ completely free.
For the resulting k-essence models, one finds that $M^2=1$, $\aB=0$, and from Eq.~(\ref{eq:cs}),
\begin{equation}
 \alpha = \aK = \frac{3\ODE(a)[1+w(a)]}{\cs^2} > 0 \,. \label{eq:alphakessence}
\end{equation}
This implies that $w(a) > -1$ given $\cs(a)^2>0$, which now becomes a free time-dependent function due to the exotic kinetic contribution.
One can furthermore restrict the model space to subluminal sound speeds $\cs(a)^2\leq1$, where quintessence is recovered in the limit of $\cs^2=1$.

Finally, note that quintessence models introduce one free function, the dark energy equation of state $-1<w(a)<1$ (or the scalar field potential) on top of the parameters of standard cosmology, where k-essence introduces a second free function in addition: the sound speed $\cs^2>0$ (or a noncanonical kinetic scalar field contribution).

\subsection{Minimally coupled cubic Galileon and Kinetic Gravity Braiding} \label{sec:KGB}

Next, we allow a nonvanishing $G_3$ term in Eq.~\eqref{eq:horndeski} while keeping $G_4=1/2$. Such a scenario is realised in the minimally coupled cubic Galileon model \cite{Nicolis2008}.
Its Lagrangian density may be written as
\begin{equation}
\mathcal{L} = \frac{R}{2} + c_2 X + 2 \frac{c_3}{H_0^2} X \Box \phi + \mathcal{L}_{\rm m}(g_{\mu\nu},\psi_{\rm m}) \,, \label{eq:cubicGalileon}
\end{equation}
with constants $c_2$, $c_3$ and a minimally coupled $\mathcal{L}_{\rm m}$.
Hence, $M^2=1$ and $\aM=0$.
It furthermore follows that $\aB=c_3 \dot{\phi}^3 H_0^{-2}H^{-1}$ and $\aK = H^{-2} \dot{\phi}^2 \left( c_2 - 12 c_3  H_0^{-2} H \dot{\phi}  \right)$, where dots indicate derivatives with respect to cosmic time here and throughout the paper.
We adopt the tracker solution $\dot{\phi} = H_0^2 H^{-1} \xi$ with constant $\xi$~\cite{DeFelice:2010pv}.
The Hubble expansion then becomes~\cite{Barreira:2014jha}:
\begin{equation}
 H^2 = \frac{1}{2}H_0^2 \left\{ \Omtoday a^{-3} + \sqrt{\Omtoday^2 a^{-6} + 4 \ODEtoday} \right\} \,.
\end{equation}
where $\ODEtoday$ represents the current fractional energy density of the Galileon field.
Note that $c_2 = 6 \xi c_3$ and $\ODEtoday = -c_3 \xi^3$.
One then finds
\begin{equation}
\alpha = 6 \ODEtoday \left(\frac{ H_0}{H}\right)^4 \left[ \ODEtoday \left(\frac{ H_0}{H}\right)^4 + 1 \right] > 0 \,,
\end{equation}
and from Eq.~\eqref{eq:cs},
\begin{equation}
 \cs^2 =  -\frac{2}{\alpha} \left[ 4\ODEtoday \frac{H_0^4 H'}{H^5} + \left( 1- \ODEtoday\frac{H_0^4}{H^4}\right) \left( \frac{H'}{H}- \ODEtoday\frac{H_0^4}{H^4}\right) + \frac{3}{2} \Om(a) \right].
\end{equation}

The cubic Galileon model introduces no additional parameter freedom over $\Lambda$CDM and makes distinct predictions for the stable EFT parametrisation $H$, $\alpha$, and $\cs^2$, particularly with a more complicated evolution of the sound speed than encountered for other examples studied here.
The model is, however, incompatible with the observed cross correlations of the integrated Sachs-Wolfe effect (ISW) in the cosmic microwave background with foreground galaxies~\cite{Lombriser2009,Kimura2011,Barreira:2014jha,Renk:2017rzu} and we do not include its expressions for the stable basis in Table~\ref{tab:modellandscape}.

The Lagrangian density~\eqref{eq:cubicGalileon} is generalised for Kinetic Gravity Braiding models to
\begin{equation}
 \mathcal{L} = \frac{R}{2} + G_2(\phi,X) + G_3(\phi,X) \Box\phi + \mathcal{L}_{\rm m}(g_{\mu\nu},\psi_{\rm m}) \,, \label{eq:KGB}
\end{equation}
which remains minimally coupled.
For example, the choice of $G_2=X$ and $G_{3\phi}=0$ with a power law of $X$ adopted for $G_{3}$ can be brought into agreement with the galaxy-ISW cross correlations for sufficiently large exponents in $G_3$~\cite{Kimura2011}, and hence provides an observationally viable cosmic acceleration without a scalar field potential or cosmological constant in $G_2$, similar to k-essence.
More generally, the cosmological phenomenology of the KGB action~\eqref{eq:KGB} to linear level is captured by the three free EFT functions $H$, $\alpha$, and $\cs^2$, where due to the minimal coupling, $M^2=1$.

\subsection{Brans-Dicke coupling with k-essence} \label{sec:JBD}

Now consider the Jordan-frame Lagrangian density $\mathcal{L} = G_4(\phi)R + G_2(\phi,X) + \mathcal{L}_{\rm m}$.
We furthermore choose the Brans-Dicke representation~\cite{Brans:1961} for the nonminimal coupling, $G_4(\phi)=\phi/2$.
Due to the absence of the $G_3$ term, it generally holds that $\aM=2\aB$.
The first derivative of the scalar field is simply given by $\dot{\phi}=H M^2\aM$, and we can use a combination of the Friedmann equations (Appendix~\ref{sec:horndeski2isb}) to eliminate $G_2$ and obtain an expression for $\ddot{\phi}$.
Using these relations in Eq.~(\ref{eq:cs}) it immediately follows that
\begin{equation}
 \cs^2=1 \,,
\end{equation}
if furthermore $G_{2XX}=0$, which is for instance the case for Jordan-Brans-Dicke gravity with a scalar field potential or for $f(R)$ gravity. 
More generally, the model space is characterised by three free EFT functions, e.g., $H^2(a)$, $M^2(a)$, and $\cs^2(a)$, with the fourth function set by the relation
\begin{equation}
 \alpha = -\frac{1}{\cs^2} \left[ \aM' + \left( \frac{H'}{H} - \frac{\aM}{2} \right)(2+\aM) + \frac{\rhom}{H^2 M^2} \right] \,, \label{eq:JBDeq}
\end{equation}
following from Eq.~\eqref{eq:cs}.

As an example, we take a Jordan-Brans-Dicke model with scalar field potential, described by $G_2=V(\phi)+X\omega(\phi)/\phi$.
This yields $\aK=\omega\aM^2$, $\alpha=(2\omega+3)\aM^2/2$, and $\cs^2=1$.
It immediately follows that the no-ghost condition implies the well-known constraint $\omega(\phi)>-3/2$.
From Eq.~(\ref{eq:JBDeq}) we obtain a second-order differential equation for $M^2=\phi$,
\begin{equation}
 \phi'' + \omega(\phi) \frac{(\phi')^2}{\phi} + \left(  \frac{H'}{H} - 1 \right) \phi' + 2\frac{H'}{H}\phi + \frac{\rhom}{H^2} = 0 \,.
\end{equation}
Due to this relation there are only two independent functions among $H(\phi)$, $\phi$, and $\omega(\phi)$, or in the EFT language, $H^2(a)$, $M^2(a)$, and $\alpha(a)$.
For a more specific example, we can further restrict to $f(R)$ gravity theories, where $\omega=0$ or $G_{2X}=G_{2XX}=0$ with $M^2=\phi=1+f_R$ and $f_R\equiv df/dR$, and hence, $\alpha=6\aB^2=3\aM^2/2$ and
\begin{equation}
 f_R'' + \left(  \frac{H'}{H} - 1 \right) f_R' + 2\frac{H'}{H}(1+f_R) + \frac{\rhom}{H^2} = 0 \,. \label{eq:fR}
\end{equation}
Here we have only one free function left, which is either $f_R=M^2-1$ or $H$.
This implies that we can either choose the $f(R)$ function to solve for $H$ or consider a designed $f(R)$ function that recovers a given $H$ in Eq.~(\ref{eq:fR}).
For an even more specific example, we can then adopt the $\Lambda$CDM expansion history to be recovered such that
\begin{equation}
 f_R'' + \left(  \frac{H'}{H} - 1 \right) f_R' + 2\frac{H'}{H}f_R = 0 \,, \label{eq:designerfR}
\end{equation}
with $H=H_{\Lambda{\rm CDM}}$.
In this case we do not have any free functions left.
The only extra freedom over $\Lambda$CDM is given by the amplitude of the growing mode of $f_R$ that is usually characterised by $f_{R0}=f_R(a=1)$.

It is furthermore convenient to define the Compton wavelength function~\cite{Song:2006ej}
\begin{equation}
 B \equiv \frac{f_R'}{1+f_R} \frac{H}{H'} = \alpha_M \frac{H}{H'} \,,
\end{equation}
and it follows that
\begin{equation}
 \alpha = \frac{3}{2} B^2 \left(\frac{H'}{H}\right)^2  = \left(\frac{3}{2}\right )^3 B^2 \Om(a)^2 \,. \label{eq:alphafR}
\end{equation}
In Sec.~\ref{sec:parametrisation}, we will use this relation together with Eq.~(\ref{eq:alphakessence}) to construct a parametrisation of $\alpha$.
Furthermore, note that from Eq.~(\ref{eq:designerfR}) we infer that at early times $f_R\propto a^{(5+\sqrt{73})/4}$ and $f_R\propto a$ in the future such that
as long as $|f_R|\ll1$, one finds $B\propto a^{(5+\sqrt{73})/4}$ and $B\propto a^4$ as well as $\alpha\propto a^{(5+\sqrt{73})/2}$ and $\alpha\propto a^2$ for early and future times, respectively.
This observation will be useful in Sec~\ref{sec:examples} for mapping $f(R)$ gravity onto a parametrisation of the stable basis as well as in Sec.~\ref{sec:approximations} for devising an approximation to the generalised version of the differential equation~\eqref{eq:designerfR}.

\subsection{Minimal self-acceleration and no-slip gravity} \label{sec:msa}

Next we shall consider a nonvanishing $G_3$ term in the action~(\ref{eq:horndeski}) with nonminimal coupling $G_4=\phi/2$ in Brans-Dicke representation.
A situation like this is encountered in the minimal modified gravity model yielding cosmic self-acceleration under the restriction that $\cT=1$~\cite{Lombriser2016a}, where $\aM=\aB$.
If dropping the relation for the minimal evolution of $M^2$ for genuine self-acceleration in Eq.~(\ref{eq:M2min}), the model generalises to no-slip gravity~\cite{Linder:2018jil}, which can further be classified in terms of partial phenomenological degeneracies between gravitational models~\cite{Lombriser2014}.
The equality between $\aM$ and $\aB$ giving rise to this terminology by setting $\gamma=1$ in Eq.~\eqref{eq:gammaQS} implies that $G_{4\phi}=XG_{3X}$ and hence that $G_3=\ln\sqrt{X}$, where the integration constant vanishes with integration by parts.

We shall first inspect the minimal self-acceleration model, for which we assume a $\Lambda$CDM expansion history $H_{\Lambda{\rm CDM}}$.
The minimal evolution in $M^2$ that is necessary to be fully responsible for driving the late-time accelerated expansion of our Universe, and to therefore genuinely attribute cosmic acceleration to a modification of gravity, is specified by~\cite{Lombriser2016a}
\begin{equation}
 \aM = \frac{C}{a H} - 2 \,,
\end{equation}
for $a\geq \aacc$ and otherwise $\aM=0$.
Furthermore, $C=2 H_0 \aacc \sqrt{3(1-\Omtoday)}$ and $\aacc=[\Omtoday/(1-\Omtoday)/2]^{1/3}$.
This yields the minimal evolution in the Planck mass
\begin{equation}
 M^2 = \left(\frac{\aacc}{a}\right)^2 e^{C(\chiacc-\chi)} \,, \label{eq:M2min}
\end{equation}
for $a\geq \aacc$ and otherwise $M^2=1$, where $\chi$ is the comoving distance $\chi(z)\equiv\int_0^z dz'/H(z')$.
The solution satisfies the no-ghost condition $M^2>0$ and we find $M_0^2\leq1$, implying an enhancement of gravity in Eq.~\eqref{eq:muQS}, which is directly linked to the aspect of self-acceleration.
One furthermore obtains
\begin{equation}
 \alpha = \frac{1}{\cs^2} \left[ \frac{2 C}{a H} - 3\left(1 + \frac{1}{M^2} \right) \Om(a) \right] \,,
\end{equation}
from Eq.~(\ref{eq:cs}).
For a choice of $\cs^2>0$, $\alpha>0$ implies
\begin{equation}
 M_0^2 > \left( \frac{2}{\sqrt{3(1-\Omtoday)}\aacc^2} - 1 \right)^{-1} \,,
\end{equation}
where $M_0^2$ corresponds to a $\sim4\%$ suppression from unity~\cite{Lombriser2016a} (see Fig.~\ref{fig:minselfacc}).
It is easy to check that this bound is satisfied for observationally viable values of $\Omtoday$.
For $a\rightarrow \aacc$, we then have $M^2\rightarrow1$ and $\alpha\rightarrow0$.
Note that the model was found to be in $3\sigma$ tension with cosmological observations, in particular due to the measured cross correlations of the ISW effect in the cosmic microwave background with foreground galaxies~\cite{Lombriser2016a}.
This implies that cosmic self-acceleration from a genuine modification of gravity in the Horndeski action is not compatible with observations given $\cT=1$.
As pointed out in Sec.~\ref{sec:EFT}, LISA Standard Sirens will improve the significance of the constraint on $M_0^2$ to the $10\sigma$ level.

The gravitational modification in Eq.~(\ref{eq:M2min}) can, however, be weakened by allowing it to interpolate to $\Lambda$CDM through adopting a rescaled value of the minimal evolution rate of the Planck mass~\cite{Lombriser:2018}
\begin{equation}
 \aM = \lambda\left(\frac{C}{a H} - 2\right) \,,
\end{equation}
with the constant $\lambda$.
Here, $\lambda>1$ allows for a sufficiently large gravitational modification for self-acceleration, $\lambda=1$ corresponds to the minimal modification,  $\lambda<1$ allows for partial self-acceleration by modified gravity, $\lambda=0$ corresponds to $\Lambda$CDM if $s=0$, and finally $\lambda<0$ implies a deceleration from modified gravity that is compensated by dark energy.
The interpolation implies $M^2=(M^2_{\rm min})^\lambda>0$ with $M^2_{\rm min}$ given by Eq.~(\ref{eq:M2min}) and
\begin{equation}
 \alpha = \frac{1}{\cs^2} \left[ \frac{2 \lambda C}{a H} - 3\left( 2\lambda - 1 + \frac{1}{M^2} \right) \Om(a) \right] \,. \label{eq:alphamsa}
\end{equation}
Models with $\lambda\geq0$ yield $\alpha\geq0$ whereas $\lambda<0$ leads to an instability due to $\alpha<0$.
If furthermore departing from a $\Lambda$CDM expansion history we can generalise this expression to
\begin{equation}
 \alpha = \frac{1}{\cs^2} \left\{ \frac{2 \lambda C}{a H} -3[1+w(a)](2\lambda-1)[ 1 - \Om(a) ] - 3\left( 2\lambda - 1 + \frac{1}{M^2} \right) \Om(a) \right\} \,.
\end{equation}
Note that we recover k-essence when $\lambda=0$, also implying $M^2=1$.
For a given $\lambda$ the generalised self-acceleration model therefore has two free EFT functions available, e.g., $H$ and $\cs^2$.

For more general no-slip gravity models we have an additional free EFT function, for instance $M^2$, where the fourth function remains specified due to the condition $\aM=\aB$.

\subsection{Horndeski models with luminal speed of gravity} \label{sec:luminalhorndeski}

Finally, the most general scalar-tensor theory in four dimensions that is Lorentz-covariant, yields at most second-order equations of motion, and is also restricted to $\cT=1$ $\forall t$ is given by the Lagrangian density~\cite{Kimura2011b,McManus2016,Creminelli:2018xsv}
\begin{equation}
 \mathcal{L} = G_4(\phi) R + G_2(\phi,X) + G_3(\phi,X)\Box\phi + \mathcal{L}_{\rm m}(g_{\mu\nu},\psi_{\rm m}) \,,\label{eq:fullLagrangian}
\end{equation}
with $G_4=\phi/2$ if adopting the Brans-Dicke representation for the modified gravity models.
In this case all of the freedom in the stable basis discussed in Sec.~\ref{sec:ISB} is needed to describe the entire available model space.
This covers all of the examples discussed in Secs.~\ref{sec:darkenergy}--\ref{sec:msa}.

In the following discussion we shall propose a parametrisation of the model space spanned by Eq.~(\ref{eq:fullLagrangian}) that allows an accurate recovery of some of the embedded examples that we have encountered here and summarised in Table~\ref{tab:modellandscape}.

\section{Parametrising the stable basis for observational tests} \label{sec:parametrising}

With the advantages of the stable basis pointed out in Sec.~\ref{sec:EFTISB}, it remains to parametrise the four free time dependent basis functions.
For this, we shall take under consideration the generality of the model space covered, its theoretical soundness, the representation of known dark energy and gravitational theories as well as simplicity and the efficiency in producing theoretical predictions for the comparison to observations.

In Sec.~\ref{sec:parametrisation} we propose and discuss a new parametrisation for the stable EFT basis.
We provide a few examples of how some typical dark energy and modified gravity models are recovered for a simple set of parameter values in Sec.~\ref{sec:examples}.
We then devise an efficient approximation to the solution of the differential equation that connects the basis to observable quantities in Sec.~\ref{sec:approximations}.
Finally, in Sec.~\ref{sec:phenomenology}, we briefly inspect some important implications regarding the general phenomenology of modified gravity models built from a stable basis.

\subsection{Parametrisation of the stable basis} \label{sec:parametrisation}

An interesting observation, made for instance in Ref.~\cite{Denissenya:2018mqs} and also in agreement with the discussion in Sec.~\ref{sec:models}, is that many modified gravity models of general interest share the property that $\aB\propto\aM$.
In the following, we shall impose this proportionality as a theoretical prior on our model space by setting
\begin{equation}
 \aM = b \, \aB \,, \label{eq:restriction}
\end{equation}
where $b$ is a constant.
Importantly, by imposing this relation we trade a free function in the basis~(\ref{eq:basis}) for a free constant.
To implement this reduction in freedom, we will remove here $M^2$ from the basis.
To study the implication of this choice, let us revisit the definition $1+\aB \equiv \tB'/\tB$ (Sec.~\ref{sec:EFTISB}), from which we find that
\begin{equation}
 M^2 = M_0^2 \left( \frac{\tB}{\tBtoday a} \right)^b \,. \label{eq:M2withb}
\end{equation}
The only condition on $\tB$ that was required in Ref.~\cite{Kennedy:2018gtx} is that it is a real quantity.
As long as $\tB$ does not change sign in its evolution, we thus have $M^2>0$, satisfying the stability requirement~\eqref{eq:stability}.
More straightforwardly, we can guarantee stability by choosing $b$ to be even.
As we have seen in Sec.~\ref{sec:JBD}, for instance, in $f(R)$ gravity we have $b=2$.
However, a change in sign of $\tB$ implies that for a smooth function, $M^2$ should vanish at some point in the evolution, which can be considered incompatible with observations, and we relax the condition on $b$ to arbitrary values.
Our new stable EFT basis under the restriction~(\ref{eq:restriction}) therefore becomes
\begin{equation}
 \quad \cs^2 > 0 \,, \quad \alpha > 0 \,, \quad \aBtoday = const. \,, \quad b=const. \,, \label{eq:basisnew}
\end{equation}
along with a free background expansion history $H$.

After these considerations, it remains to specify the other three free functions of the basis~\eqref{eq:basisnew}, and we propose here the parametrisation
\begin{eqnarray}
 H & : & \quad w(a) = w_0 + (1-a) w_a \,, \label{eq:w0wa} \\
 M^2 & : & \quad \aM(a) = b \: \aB(a) \,, \\
 \cs^2 & : & \quad \cs^2(a) = 1 + s \frac{\ODE(a)}{\ODEtoday} \,, \label{eq:cs2param} \\
 \alpha & : & \quad \alpha(a) = \frac{\alpha_0 (1+s) + \alpha_1 (1-a^{u_1}) }{\cs^2} \left( \frac{\ODE(a)}{\ODEtoday} \right)^{u_2} \,, \label{eq:alphaparam}
\end{eqnarray}
where we have the ten free parameters $w_0$, $w_a$, $M_0^2$, $b$, $\aBtoday$, $s$, $\alpha_0$, $\alpha_1$, $u_1$, and $u_2$.
We emphasise that not all of the ten parameters are equally important in constraining or reporting a deviation from concordance cosmology.
Particularly, the four parameters introduced in $\alpha$ can typically be marginalised over.
Specifically, $\Lambda$CDM is recovered for $w_0=-1$, $w_a=0$, $M_0^2=1$, $\aBtoday=0$, arbitrary $b$, $s=0$, $\alpha_0=0$, and at least either $u_1=0$ or $\alpha_1=0$.
Hence, evidence for modified gravity requires $M_0^2\neq1$, for dynamical dark energy $w_0\neq-1$ or $w_a\neq0$, and for more exotic dark energy models $\aBtoday\neq0$ or $s\neq0$.
We provide some specific examples in Sec.~\ref{sec:examples}.

We adopt the Chevallier-Polarski-Linder (CPL)~\cite{Chevallier:2000qy,Linder:2002et} parametrisation of the equation of state $w(a)$ in Eq.~\eqref{eq:w0wa} due its frequent use.
It also conveniently implies that $H^2\ODE(a) = H_0^2\ODEtoday a^{-3(1+w_0+w_a)} e^{3w_a(a-1)}$ such that there is no need of integrating the dark energy conservation equation.
For $\cs^2$ we chose a simple parametrisation in Eq.~\eqref{eq:cs2param} that traces the dominant energy component and may easily be further generalised if needed.
The parametrisation in Eq.~(\ref{eq:alphaparam}) is constructed such that it reproduces quintessence and k-essence models exactly (Sec.~\ref{sec:darkenergy}) and also accurately recovers the evolution of the Compton wavelength function in $f(R)$ gravity (Sec.~\ref{sec:JBD}) (also see Sec.~\ref{sec:examples}).
The sound speed is used in the denominator due to the relevance of the combination of $\alpha\cs^2$.
The number of parameters used in Eq.~(\ref{eq:alphaparam}) is minimal: the three parameters $\alpha_1$, $u_1$, and $u_2$ describe the broken power-law evolution encountered in $f(R)$ gravity whereas $\alpha_0$ is required for the exact recovery of quintessence and k-essence models.
Note that it is important to achieve good accuracy for $\alpha\cs^2$ as it is used to solve for $M^2$ or $\aB$ in Eqs.~\eqref{eq:diffeq} or \eqref{eq:aBdiffeq}.
Due to the no-ghost condition $\alpha>0$ and the requirement that $\alpha\rightarrow0$ at early times, one finds that if $u_1>0$, we must have $\alpha_1<0$ and either $\alpha_0(1+s)>-\alpha_1$ with $u_2>0$ or $\alpha_0(1+s)=-\alpha_1$ with $u_1-3 (w_0+w_a) u_2>0$.
Note that $w_0 + w_a<0$ for matter domination in the past.
If $u_1=0$ or $\alpha_1=0$, we require $\alpha_0>0$ and $u_2>0$.
Finally, if $u_1<0$, we must have $\alpha_0(1+s)\geq-\alpha_1$ and $\alpha_1<0$ with $u_1-3 (w_0+w_a) u_2>0$,
It is also required that $s>-1$ to obtain $\cs^2>0$ and avoid gradient instabilities, and $s\leq0$ if requiring $\cs^2\leq1$.
Note that we cannot have $\aM\neq0$ and $\aB=0$, which corresponds to $G_3 = \ln X^{-1/2}$ for a Brans-Dicke coupling.
However, it holds that $G_3 = \ln X^{1/b-1/2}$, which hence approximates this solution in the limit of $b\rightarrow\infty$.

In order to connect the modifications introduced with Eqs.~\eqref{eq:w0wa}--\eqref{eq:alphaparam} to the $\alpha_i$ functions or the phenomenological modifications $\mu$ and $\gamma$ in Sec.~\ref{sec:EFT}, we need to solve a differential equation for $M^2$ or $\aB$.
But we shall discuss in Sec.~\ref{sec:approximations} how one may evade the integration of these equations.
When $b$ is free, generally sampling $b\neq0$, we solve the relation
\begin{eqnarray}
 (M^2)'' + \frac{1-2b}{b M^2} [(M^2)']^2 - \frac{1+2b+3\ODE(a) w(a)}{2} (M^2)'  & & \nonumber\\
  + \frac{b \{ 3 - 3\ODE(a) + M^2[\alpha\cs^2 - 3 - 3\ODE(a) w(a) ] \} }{2} & = & 0 \,, \label{eq:diffeq}
\end{eqnarray}
which follows from Eq.~\eqref{eq:cs} with boundary conditions $M^2_0$ and $(M^2)'|_0 = M^2_0 \aMtoday= b M^2_0 \aBtoday$.

In the case of having exactly $b=0$, for example, from a limitation of the parameter space to exotic dark energy models excluding modified gravity, we may instead set $M^2=1$ and solve for
\begin{equation}
 \aB' + \aB^2 - \frac{1+3\ODE(a) w(a)}{2}\aB + \frac{\alpha \cs^2 - 3\ODE(a)[1+w(a)]}{2} = 0 \label{eq:aBdiffeq}
\end{equation}
with the boundary condition given by the parameter $\aBtoday$.
Using the substitution $1+\aB \equiv \tB'/\tB$ defined in Sec.~\ref{sec:EFTISB} we derive the ordinary homogeneous second-order differential equation
\begin{equation}
 \tB'' - \frac{5+3\ODE(a) w(a)}{2}\tB' + \frac{\alpha \cs^2 + 3\Om(a)}{2} \tB = 0 \,.
\end{equation}
Note that for k-essence, one finds $\aB=0$.

\subsection{Examples} \label{sec:examples}

To study how the parametrisation~\eqref{eq:w0wa}--\eqref{eq:alphaparam} performs in practice, let us now inspect the parameter values for four example models: quintessence, k-essence, designer $f(R)$ gravity, and generalised minimal self-acceleration.
We summarise the results in Table~\ref{tab:models}.

\begin{table}
    \centering
    \begin{tabular}{l|c|c|c|c|c}
      & $\Lambda$CDM & quintessence & k-essence & designer $f(R)$ & gen-min/no-$\gamma$ \\
     \hline
     $w_0$ & $-1$ & $w_0$ & $w_0$ & $-1$ & $-1$ \\
     $w_a$ & 0 & $w_a$ & $w_a$ & 0 & 0 \\
     $M_0^2$ & 1 & 1 & 1 & $1+f_{R0}$ & $\aacc^{2\lambda} e^{\lambda C(\chiacc-\chi_0)}$ \\
     $b$ & - & - & - & 2 & 1 \\
     $\aBtoday$ & 0 & 0 & 0 & $-\frac{3}{4}\Omtoday B_0$ & $\lambda\left(\frac{C}{H_0}-2\right)$ \\
     $s$ & 0 & 0 & $s$ & 0 & $s$ \\
     $\alpha_0$ & 0 & $3\ODEtoday(1+w_0)$ & $\frac{3\ODEtoday(1+w_0)}{1+s}$ & $6\aBtoday^2$ & $-\alpha_1(1+s)^{-1}$ \\
     $\alpha_1$ & 0/- & $3\ODEtoday w_a$ & $3\ODEtoday w_a$ & $-\alpha_0$ & $2\lambda (3\Omtoday - C H_0^{-1})$ \\
     $u_1$ & -/0 & 1 & 1 & 2 & $-1$ \\
     $u_2$ & - & 1 & 1 & $\frac{1}{6}(1+\sqrt{73})$ & $1 + \frac{\aBtoday-2\lambda}{\alpha_1}$
    \end{tabular}
    \caption{Parameter values for the parametrised stable EFT basis~\eqref{eq:w0wa}--\eqref{eq:alphaparam} for $\Lambda$CDM and a few frequently studied dark energy and modified gravity models.
    The designer $f(R)$ gravity model adopts a $\Lambda$CDM expansion history, and one can use the approximation $B_0=-2.1\Om^{-0.76}f_{R0}$~\cite{Ferraro:2010} to relate the Compton wavelength parameter $B_0$ to $f_{R0}$.
    The generalised minimal self-acceleration model (gen-min) rescales the minimal evolution in $M^2$ required for self-acceleration with an exponent $\lambda$ and represents a no-slip gravity (no-$\gamma$) scenario, where $G_3=\ln\sqrt{X}$ in Brans-Dicke representation.
    }
    \label{tab:models}
\end{table}

\paragraph{Quintessence/k-essence:} As discussed in Sec.~\ref{sec:darkenergy}, quintessence models are specified by just one free function: their scalar field potential $V(\phi)$.
The respective stable EFT basis is given by $M^2=\cs^2=1$, and $\aBtoday=0$, as in $\Lambda$CDM, where the basis functions that deviate from standard cosmology are $\alpha=\aK$ and $H$.
Importantly, the two deviating functions are related since we only have one underlying free function characterising the model space.
We have seen in Sec.~\ref{sec:darkenergy} that $\alpha=3[1-\Om(a)][1+w(a)]=3\ODE(a)[1+w(a)]$.
Hence we can simply define $w(a)>-1$ as the free function of the model, which defines $H$ and $\alpha>0$.
But we could inversely have picked $\alpha>0$ to solve for $w(a)$ (and $H$) instead.
For the parametrisation~\eqref{eq:w0wa}--\eqref{eq:alphaparam} this implies that the parameter values $w_0$, $w_a$ and the ones in $\alpha$ need to be consistent.
More specifically, we have $w_0 > -1$, $w_a > -1-w_0$, $M_0^2=1$, arbitrary $b$, $\aBtoday=0$, $s=0$, $\alpha_0=3\ODEtoday(1+w_0)$, $\alpha_1=3\ODEtoday w_a$, and $u_1=u_2=1$.
Note that this choice of parameters recovers the expressions of quintessence in the stable EFT basis exactly.

These results can easily be generalised to k-essence, taking $-1<s\neq0$, which implies $\alpha_0=3\ODEtoday(1+w_0)/(1+s)$, where the other parameter values are as in quintessence.
This also reproduces the expressions found in Sec.~\ref{sec:darkenergy} exactly.

\paragraph{$f(R)$ gravity:}
For another example we revisit the designer $f(R)$ gravity model with $\Lambda$CDM expansion history and
inspect its recovery from adopting the parametrisation~\eqref{eq:w0wa}--\eqref{eq:alphaparam} for the stable basis.
The model introduces only one free parameter over $\Lambda$CDM: $f_{R0}$ or $B_0$.
Hence, the parametrisation of the model should only contain this freedom.
From the calculations in Sec.~\ref{sec:JBD}, we immediately identify $w_0=-1$, $w_a=0$, $M_0^2=1+f_{R0}$, $b=2$, and $s=0$.
Furthermore, we have
\begin{equation}
 \aBtoday \equiv \frac{1}{2}\frac{f_{R}'|_0}{1+f_{R0}} = - \frac{3}{4} \Omtoday B_0 \,,
\end{equation}
where $B_0$ can be related to $f_{R0}$ by the approximation $B_0=-2.1\Om^{-0.76}f_{R0}$~\cite{Ferraro:2010}.
Finally, we have the stability function $\alpha=6\aB^2=3\aM^2/2=3[(M^2)'/M^2]^2/2$.
In principle, we could now use this relation along with the other specified parameter expressions in Eq.~\eqref{eq:diffeq} to directly recover the $f_R$ field equation~\eqref{eq:designerfR}.
Here we shall instead recover the $f_R$ field from the parametrisation~\eqref{eq:w0wa}--\eqref{eq:alphaparam}.

\begin{figure}
 \centering
 \resizebox{0.496\textwidth}{!}{
 \includegraphics{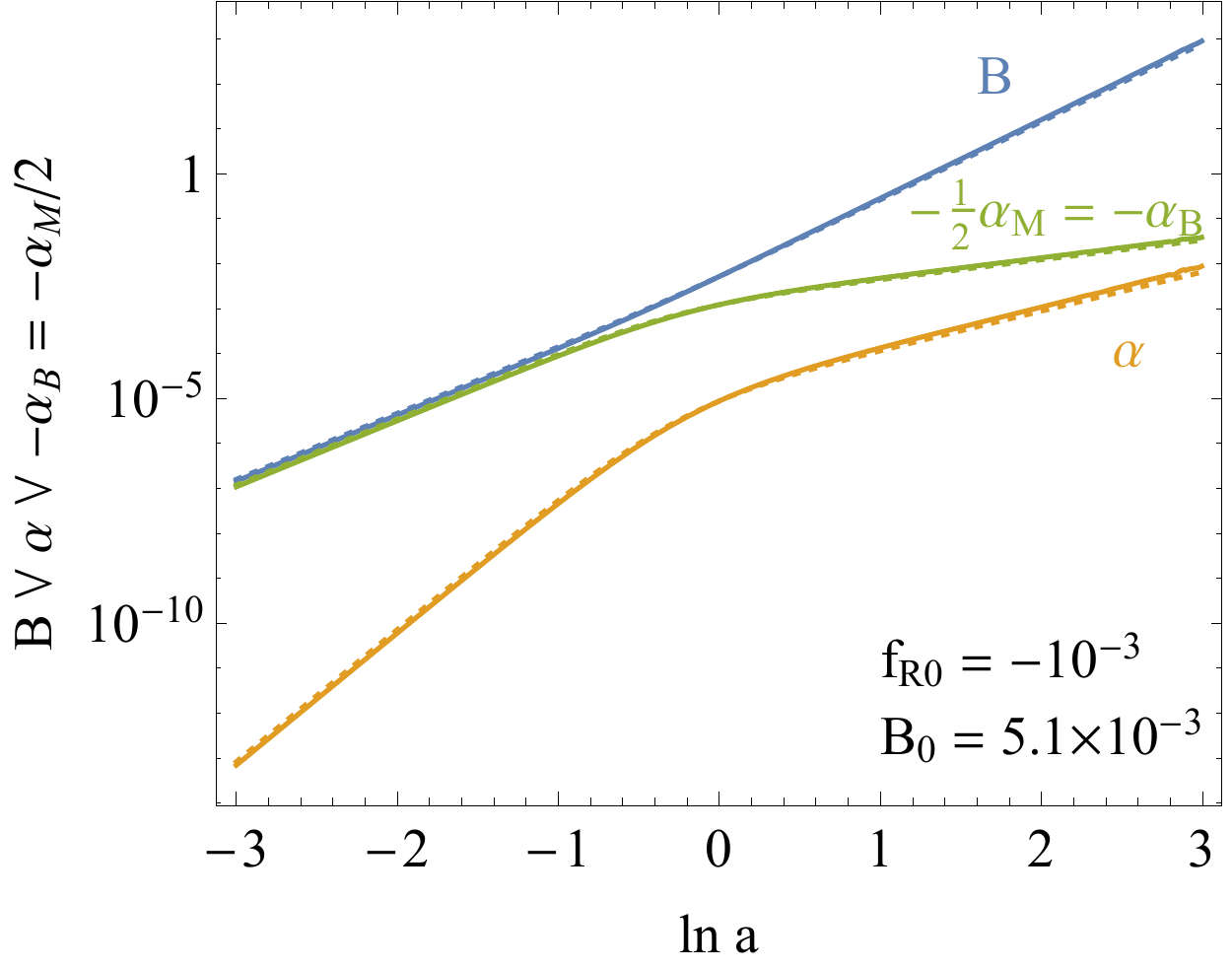}
 }
 \resizebox{0.496\textwidth}{!}{
 \includegraphics{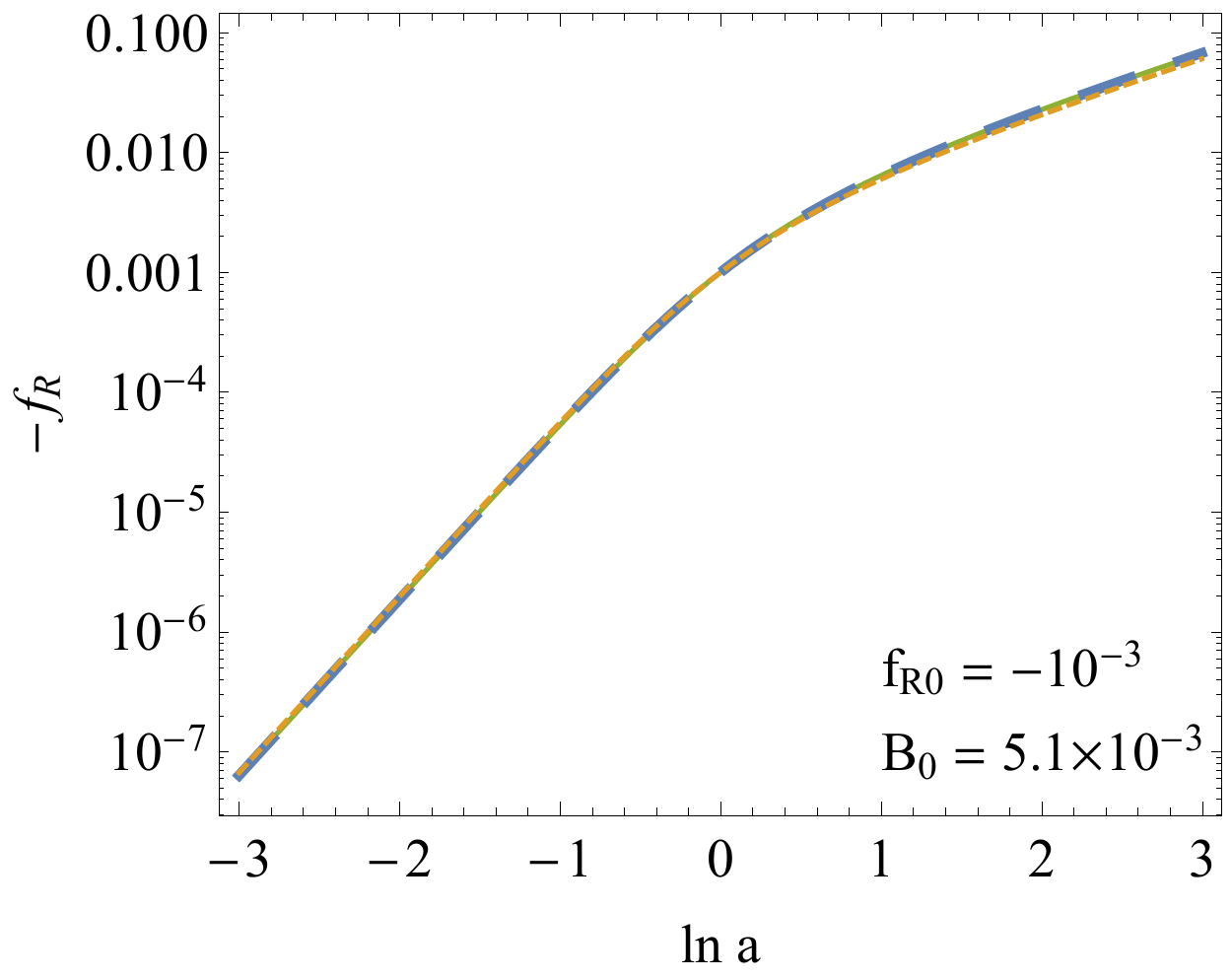}
 }
\caption{
Recovery of $f(R)$ gravity in the parametrisation of the stable EFT basis by Eqs.~\eqref{eq:w0wa}--\eqref{eq:alphaparam}.
Results are shown adopting Planck 2018 cosmological parameters~\cite{Aghanim:2018eyx}.
\emph{Left panel:} The solid curves show the exact numerical results evaluated with the code of Refs.~\cite{Lombriser:2010mp,Lombriser:2011zw} whereas the dotted curves illustrate the parametrised counterparts with the constants $u_1$ and $u_2$ in $\alpha(a)$ determined by the early and future limits of the Compton wavelength function $B(a)$.
One can avoid the integration of Eq.~\eqref{eq:diffeq} by adopting the approximation for $\aB(a)$ given by  $B(a)$.
\emph{Right panel:} The solid curve shows the exact numerically evaluated $f_R$ function, the long-dashed curve corresponds to the counterpart obtained from the integration of Eq.~\eqref{eq:diffeq} with the parametrisation of $\alpha$ in Eq.~\eqref{eq:alphaparam}, and the short-dashed curve corresponds to the approximation~\eqref{eq:fRapprox} that avoids the integration.
}
\label{fig:fRgravity}
\end{figure}

Specifically, we need to map the model onto $\alpha$ in Eq.~\eqref{eq:alphaparam}, which is then used in Eq.~\eqref{eq:diffeq} along with the other parameter values to solve for $M^2$.
To accomplish this, we make use of the power-law behaviour of $B(a)$ at early and future times, the large positive and negative redshift limits in the deep matter and dark energy dominated regimes, respectively, but where $|f_R|\ll1$ as discussed in Sec.~\ref{sec:JBD}.
We then employ powers of $\Om(a)$ or $\ODE(a)$ to naturally transition between the two regimes.
More specifically, we make the ansatz $B(a)\propto[\ODE(a)]^m a^n$.
Note that we could equivalently use $\Om(a)$ instead, as $\Om(a)\propto\ODE(a)a^{-3}$ for $H_{\Lambda{\rm CDM}}$.
The exponents $m$ and $n$ are solved by the two equations obtained for the power-law behaviour of $B(a)$ in the early and future limits of Eq.~\eqref{eq:diffeq} for a regime where $M^2\approx1$.
With this procedure we find the approximation
\begin{equation}
 B \approx B_0 \left[ \frac{\ODE(a)}{\ODEtoday} \right]^{\frac{-11+\sqrt{73}}{12}}a^4 \,, \label{eq:Bapprox}
\end{equation}
which we compare against the numerical result produced by the code of Refs.~\cite{Lombriser:2010mp,Lombriser:2011zw} in Fig.~\ref{fig:fRgravity}, finding very good agreement with the exact solution.
Similarly, we find that $\aB = -3\Om B/4 = \aM/2$ is well reproduced (see Fig.~\ref{fig:fRgravity}).
Hence, from Eq.~\eqref{eq:alphafR} it then follows that
\begin{equation}
 \alpha \approx \left(\frac{3}{2}\right)^3 \Omtoday^2 B_0^2 \left[ \frac{\ODE(a)}{\ODEtoday} \right]^{\frac{1+\sqrt{73}}{6}}a^2 \,, \label{eq:alphaapprox}
\end{equation}
which can easily be mapped onto Eq.~\eqref{eq:alphaparam} (see Table~\ref{tab:models}) and accurately recovers the exact numerical solution (Fig.~\ref{fig:fRgravity}).

The procedure intended for the parametrisation~\eqref{eq:w0wa}--\eqref{eq:alphaparam} of the stable basis then requires the integration of Eq.~\eqref{eq:diffeq} with the $\alpha$ given by Eq.~\eqref{eq:alphaapprox} to determine $M^2$, or $f_R$ in this case.
Performing this integration we find excellent agreement with the numerical solutions (Fig.~\ref{fig:fRgravity}).
Importantly, the fact that we have to solve a differential equation for $f_R$ should not be seen as a flaw of the approach, but rather it shows that the parametrisation is physically well grounded.
We would not expect any different from studying the designer $f(R)$ model directly, i.e., we would have to solve for $f_R$ with Eq.~\eqref{eq:fR} rather than parametrising the time dependence of the field directly.
As we have seen in Sec.~\ref{sec:JBD} a parametrisation of $f_R$ would require solving a differential equation for the expansion history $H$ with Eq.~\eqref{eq:fR} instead.

However, for reasons of efficiency we may choose to avoid this integration and work with an approximation of $f_R$ given by the integration of Eq.~\eqref{eq:Bapprox}.
For designer $f(R)$ gravity, we find that
\begin{equation}
 f_R \approx f_{R0} \left[ \frac{\ODE(a)}{\ODEtoday} \right]^{\frac{-11+\sqrt{73}}{12}}a^4 \frac{{}_2F_1\left[-1,\frac{4}{3};\frac{17+\sqrt{73}}{12};-\frac{\ODE(a)}{\Om(a)}\right]}{{}_2F_1\left[-1,\frac{4}{3};\frac{17+\sqrt{73}}{12};-\frac{\ODEtoday}{\Omtoday}\right]} \,, \label{eq:fRapprox}
\end{equation}
where ${}_2F_1$ denotes the hypergeometric function
\begin{equation}
 {}_2F_1(a,b;c;z) = \sum_{n=0}^{\infty} \frac{(a)_n(b)_n}{(c)_n} \frac{z^n}{n!}
\end{equation}
with rising Pochhammer symbols
\begin{equation}
 (q)_n = \left\{ \begin{array}{ll} 1 \,, & n = 0 \\ q(q+1)\cdots(q+n-1) \,, & n>0 \end{array} \right. \,.
\end{equation}
In Fig.~\ref{fig:fRgravity}, we compare the approximation in Eq.~\eqref{eq:fRapprox} against numerical results obtained from employing the code of Refs.~\cite{Lombriser:2010mp,Lombriser:2011zw} and find excellent agreement between the two.
Note that we may further approximate the hypergeometric functions in Eq.~\eqref{eq:fRapprox} with an expression like $[\ODE(a)]^m a^n$ or similarly to $B$, obtain the exponents directly from the early and future power-law behaviours of $f_R$ such that $f_R \approx f_{R0} [\ODE(a)/\ODEtoday]^{(1+\sqrt{73})/12}a$, which is simpler however less accurate than Eq.~\eqref{eq:fRapprox} but can easily be generalised (Sec.~\ref{sec:approximations}).

\paragraph{Minimal self-acceleration}
Recall that for nonvanishing $G_{3X}$ we have $b\neq2$, and if furthermore $\aM$ is nonvanishing, it follows that $G_{3X} = \ln X^{1/b-1/2}$ where the integration constant vanishes with integration by parts.
As specified before, the case where $b=1$ corresponds to the minimal self-acceleration or no-slip gravity scenario.
For the extended minimal self-acceleration model (Sec.~\ref{sec:msa}), still adopting a $\Lambda$CDM expansion history, we find that
\begin{equation}
 \frac{\alpha\cs^2}{2\lambda} = \frac{C}{a H} - 3 \Om(a) (1+\epsilon) = \bar{C} \left(\frac{\ODE(a)}{\ODEtoday}\right)^{1/2}a^{-1} - 3\Omtoday  \left(\frac{\ODE(a)}{\ODEtoday}\right) a^{-3} (1+\epsilon) \,, \label{eq:acs}
\end{equation}
for $a>\aacc$ and $\alpha=0$ for $a\leq\aacc$, where we defined the function $\epsilon\equiv(M^{-2}-1)/(2\lambda)$ and the parameter $\bar{C} \equiv H_0^{-1} C$.
Following the same procedure to map Eq.~\eqref{eq:acs} onto the parametrisation~\eqref{eq:alphaparam} as for $f(R)$ gravity, we can find from the future limit that $u_1=-1$ and $\alpha_1=-\alpha_0(1+s)$ as long as $\lambda\leq1$, otherwise $u_1=-3+2\lambda$.
This follows from the behaviour of $\epsilon\propto a^{2\lambda}$ in the future and $\epsilon\rightarrow0$ in the past, which can be inferred from Eq.~\eqref{eq:M2min}.
We can then match the parametrisation at $\alpha(a=1)$, which implies that $\alpha_0=2\lambda[\bar{C}-3\Omtoday(1+\epsilon_0)]/(1+s)$.
As $\alpha=0$ in the far past by construction, we then must have $u_2>-u_1/3$.
The $\alpha$ function is not smooth at $\aacc$, which is why we loose information on $u_2$.
To introduce an additional constraint, we shall require that $(\alpha\cs^2)'$ is matched at the present, which gives
\begin{equation}
 u_2 = 1+ \frac{\aBtoday - 2\lambda \left[ 1 + (1+u_1)\left( 1 + \epsilon_0 - \frac{\bar{C}}{3\Omtoday} \right) + 2\epsilon_0 - \epsilon_0' \right]}{\alpha_1} \,. \label{eq:u2}
\end{equation}
This simplifies to $u_2=1 + (\aBtoday-2\lambda)/\alpha_1$ for $\lambda\leq1$ and neglecting contributions of $\epsilon$, which are the values we quote in Table~\ref{tab:models}.
Including the $\epsilon$ contributions, we note that $2\lambda\epsilon_0 = M_0^{-2}-1$ and $2\lambda\epsilon_0' = - M_0^{-2}\aMtoday$.
$M_0^2$ can easily be evaluated with Eq.~\eqref{eq:M2min} and $\aMtoday=\lambda(\bar{C}-2)$.
For $H=H_{\Lambda{\rm CDM}}$ we can use that $\chi(a) = {}_2F_1[1/3,1/2;4/3;-\Om(a)/\ODE(a)]/(a\sqrt{\ODEtoday})$ to obtain $M^2(a)$ and $M^2_0$.
Similarly to $f(R)$ gravity $M^2$ can be approximated directly with a combination of powers of $\ODE(a)$ and $a$ to avoid the integration of Eq.~\eqref{eq:diffeq} (Sec.~\ref{sec:approximations}).

\begin{figure}
 \centering
 \resizebox{0.496\textwidth}{!}{
 \includegraphics{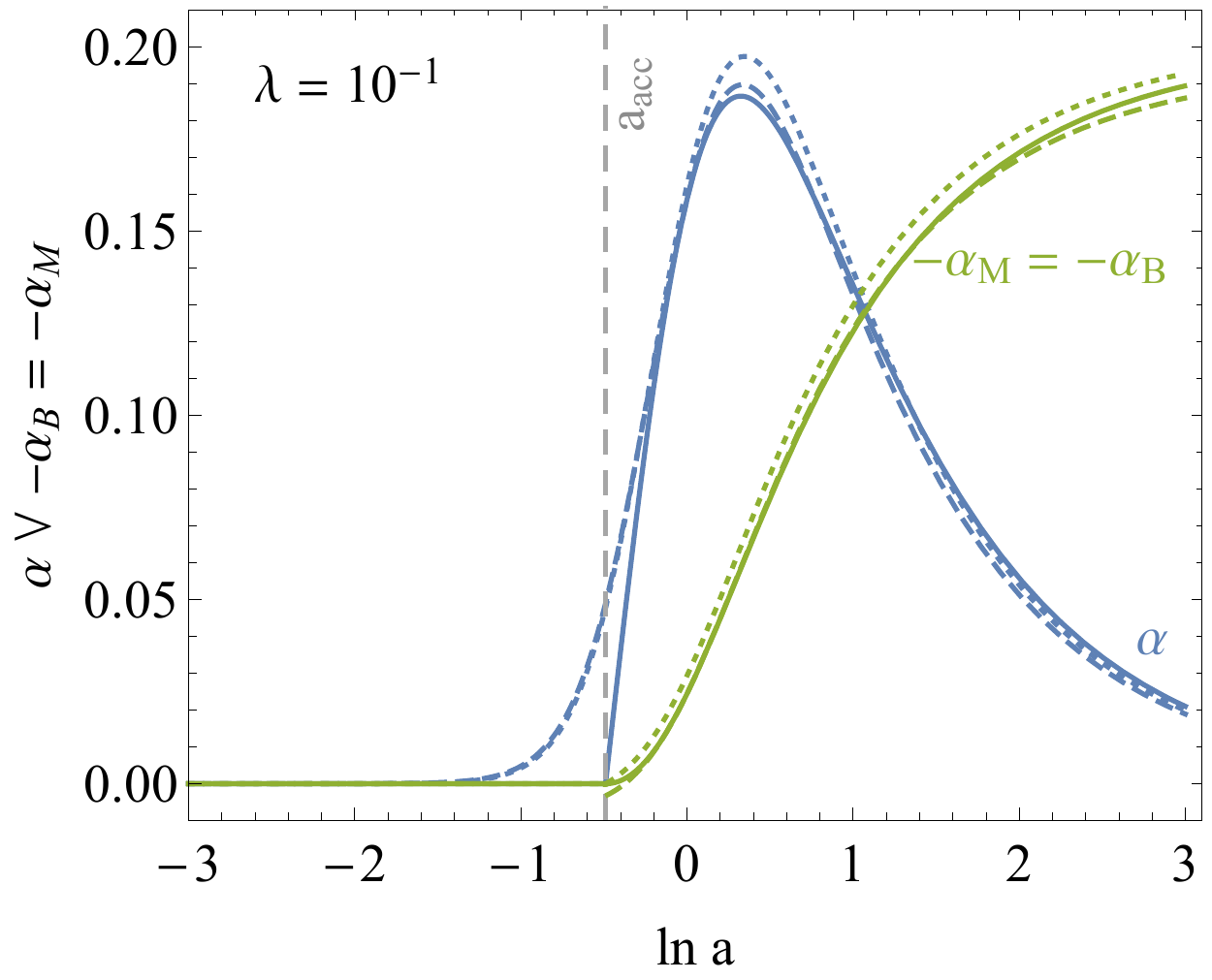}
 }
 \resizebox{0.496\textwidth}{!}{
 \includegraphics{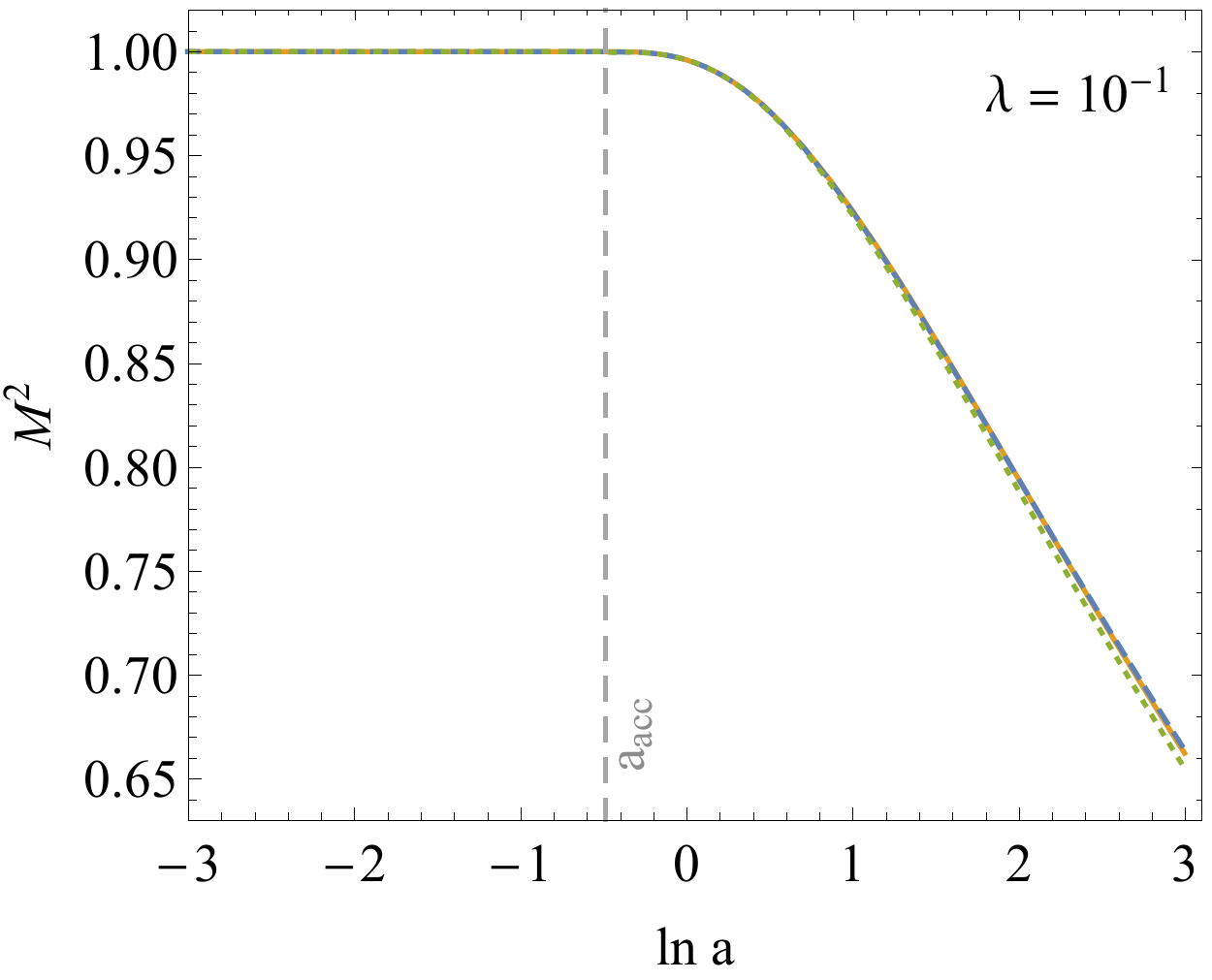}
 }
\caption{Recovery of the minimal self-accelerated modification of gravity from a parametrisation in the stable EFT basis~\eqref{eq:w0wa}--\eqref{eq:alphaparam}.
Solid curves correspond to the exact numerical solutions for the model whereas the other curves represent the recovery of the model in the parametrisation~\eqref{eq:w0wa}--\eqref{eq:alphaparam}.
The dashed curves illustrate results from employing an approximation that includes the deviation in the Planck mass and the dotted curve illustrates the scenario where this deviation has been neglected.
}
\label{fig:minselfacc}
\end{figure}

In Fig.~\ref{fig:minselfacc} we compare the parametrised $\alpha$, both when including and not including the contributions of $\epsilon_0$ and $\epsilon_0'$ in Eq.~\eqref{eq:u2}, against the exact result from Eq.~\eqref{eq:acs}, where we set $\lambda=10^{-1}$, $s=0$, and adopt Planck cosmological parameters~\cite{Aghanim:2018eyx}.
We furthermore show the Planck mass and its evolution rate that follow from the integration of these approximations with Eq.~\eqref{eq:diffeq} against the exact expressions.
Note that we solve the differential equation only in the domain of $a\geq\aacc$.

\subsection{Efficient approximations} \label{sec:approximations}

The parametrisation adopted for the stable EFT basis with Eqs.~\eqref{eq:w0wa}--\eqref{eq:alphaparam} requires solving a differential equation for $M^2$ or $\aB$, Eqs.~\eqref{eq:diffeq} or \eqref{eq:aBdiffeq}, to fully determine the $\alpha_i$ functions and take advantage of existing numerical linear Boltzmann solvers~\cite{Hu:2013twa,Zumalacarregui:2016pph}.
This is generally not faster than simply parametrising the $\alpha_i$ functions and sampling for stable parameter regions~\cite{Denissenya:2018mqs} (Sec.~\ref{sec:EFTISB}).
However, as mentioned in Sec.~\ref{sec:examples}, the need for solving a differential equation is not unexpected as one also encounters this when studying specific modified gravity models such as $f(R)$ gravity or Jordan-Brans-Dicke theories more generally (Sec.~\ref{sec:JBD}).
This may thus as well be perceived as a sign of performing a more physical parametrisation of the model space than by, for instance, choosing simple power laws for the $\alpha_i$.
Yet, ideally we would be able to analytically integrate the differential equation or provide an accurate approximation for its solution to improve the numerical efficiency.
In the following we shall briefly explore how a simple but general approximation for the evolution of the Planck mass $M^2$ can be inferred for the parametrisation~\eqref{eq:w0wa}--\eqref{eq:alphaparam} from Eq.~\eqref{eq:diffeq}.

Generally, the modifications around $\Lambda$CDM allowed by observational data are small.
For example, the minimal self-accelerated modification of gravity (Sec.~\ref{sec:msa}) requires a suppression of $M_0^2$ of about 4\% from unity, and this was found to be in $3\sigma$ tension with cosmological observations~\cite{Lombriser2016a} (Sec.~\ref{sec:msa}).
It can therefore generally be expected that observationally viable Horndeski modifications of gravity should be limited to percent-level deviations from $M_0^2=1$.
For our approximation, we shall therefore assume that $M^2\approx1$ in Eq.~\eqref{eq:diffeq}, which typically remains accurate for about an $e$-folding of the scale factor into the future (see Figs.~\ref{fig:fRgravity} and \ref{fig:minselfacc}).

As we consider a universe of matter domination in the past and dark energy domination in the future, we can simply assume that the evolution of the Planck mass follows a broken power law in the scale factor with future and past limits $(M^2-1)_{f/p}=A_{f/p} a^{f/p}$, where $f$ and $p$ denote the exponents in the corresponding limits and $A_{f/p}$ are constant amplitudes.
In the following we shall not present a full case-by-case study but provide a few examples for how a generalised approximation for the evolution of the Planck mass can be obtained.
For simplicity we shall furthermore restrict the discussion to $w_a=0$.

Let us first inspect the case where $u_1>0$ and $\alpha_1<0$ and extract the future limit $(\alpha\cs^2)_f$ from the parametrisation~\eqref{eq:alphaparam}.
In order to have a counter term that cancels $(\alpha\cs^2)_f$ in the future limit of Eq.~\eqref{eq:diffeq}, we must have $f>0$.
The dominating terms then become
\begin{equation}
 \frac{(4b-2)f^2}{b^2} A_f^2 a^{2f} = (\alpha\cs^2)_f \,, \label{eq:futdom1}
\end{equation}
from which we immediately identify that $f=u_1/2$.
We discard the information on the amplitude $A_f$ as we are only interested in the evolution rate.
For the past limit, we find that the dominating terms for $p>0$ satisfy the relation
\begin{equation}
 p^2 - \frac{1+2b}{2}p - \frac{3b}{2} \left[ 1 - \frac{(\alpha\cs^2)_p}{3 A_p a^p} \right] = 0 \label{eq:pastdom}
\end{equation}
and hence for $(\alpha\cs^2)_p \ll A_p a^p$, one finds that $p = (1+2b+\sqrt{1+28b+4b^2})/4$ as long as $b\geq(4\sqrt{3}-7)/2$.

Next, we consider the case $u_1<0$ with $\alpha_0(1+s)=-\alpha_1$ and $w_0=-1$.
For the future limit in Eq.~\eqref{eq:diffeq} one finds that if $u_1=2(b-1)$ then the dominant terms are again given by Eq.~\eqref{eq:futdom1} with $f=u_1/2$.
Otherwise, the dominating terms become
\begin{equation}
 \frac{2f(f+1-b)}{b}A_f a^f = (\alpha\cs^2)_f \,, \label{eq:futdom2}
\end{equation}
which yields $f=u_1$.
For the past limit, we again find Eq.~\eqref{eq:pastdom} with the quadratic solution for $p$ unless  $(\alpha\cs^2)_p \centernot\ll A_p a^p$, in which case $p=3u_2+u_1$.

Let us now revisit the designer $f(R)$ gravity model, where $u_1>0$ and $\alpha_1<0$.
From the parameter values in Table~\ref{tab:models} we find that $f=1$ and $p=(5+\sqrt{73})/4$.
For the extended minimal self-acceleration model studied in Fig.~\ref{fig:minselfacc}, where $u_1<0$, $\alpha_0(1+s)=-\alpha_1$, and $w_0 =-1$, we obtain $f=-1$ and $p=(\bar{C}+12\ODEtoday)/(2\bar{C}-6\Omtoday)$.
As discussed in Sec.~\ref{sec:examples}, a well-motivated transition function connecting the past and future power-law behaviours of $M^2$ is simply given by the fractional energy density $\ODE(a)$ or $\Om(a)$.
Specifically, we shall make the ansatz
\begin{equation}
 M^2= 1+ (M^2_0 -1) \left[\frac{\ODE(a)}{\ODEtoday} \right]^m a^n \label{eq:M2approx}
\end{equation}
with constants $m$ and $n$.
One generally finds that $n=f$ and $m = (f-p)/(3w_0)$.
For designer $f(R)$ gravity this implies that $n=1$ and $m=(1+\sqrt{73})/12$, and for the minimal self-acceleration model of Fig.~\ref{fig:minselfacc}, the exponents are given by $n=-1$ and $m=(4+\bar{C}-6\Omtoday)/(2\bar{C}-6\Omtoday)$.

Alternatively to fixing $m$ and $n$ with the future and past limits of Eq.~\eqref{eq:diffeq} and $\alpha\cs^2$, we can replace one constraint with a match of the current evolution rate of $M^2$ in Eq.~\eqref{eq:M2approx} to $\aMtoday=b\aBtoday$.
This yields the relation
\begin{equation}
 m = \frac{1}{3\Omtoday w_0} \left(n + \frac{b \aBtoday M^2_0}{1-M^2_0} \right) \,, \label{eq:malt}
\end{equation}
which in combination with a constraint from $f$ or $p$ fixes $m$ and $n$.
For simplicity, we shall adopt here $n=f$.

\begin{figure}
 \centering
 \resizebox{0.5\textwidth}{!}{
 \includegraphics{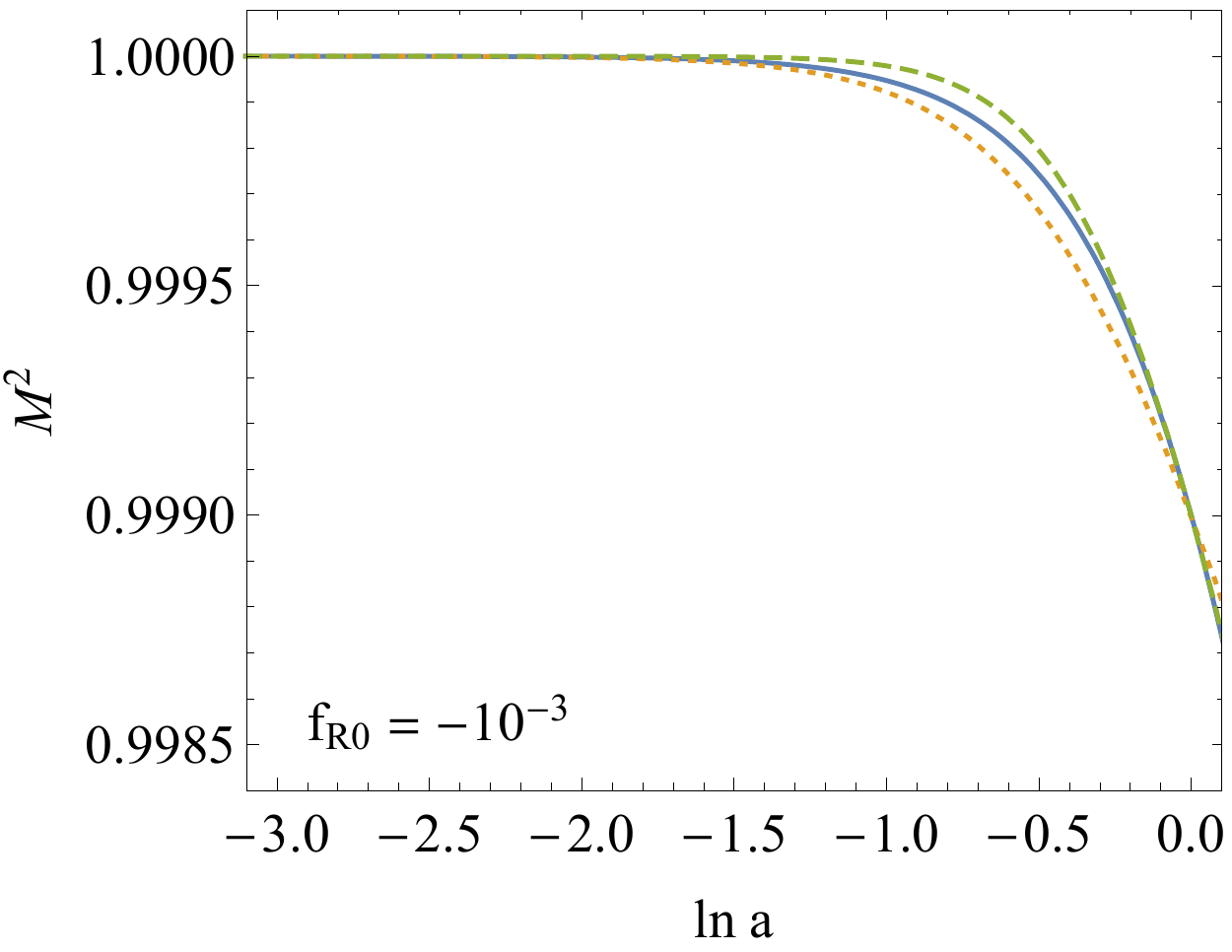}
 }
 \resizebox{0.488\textwidth}{!}{
 \includegraphics{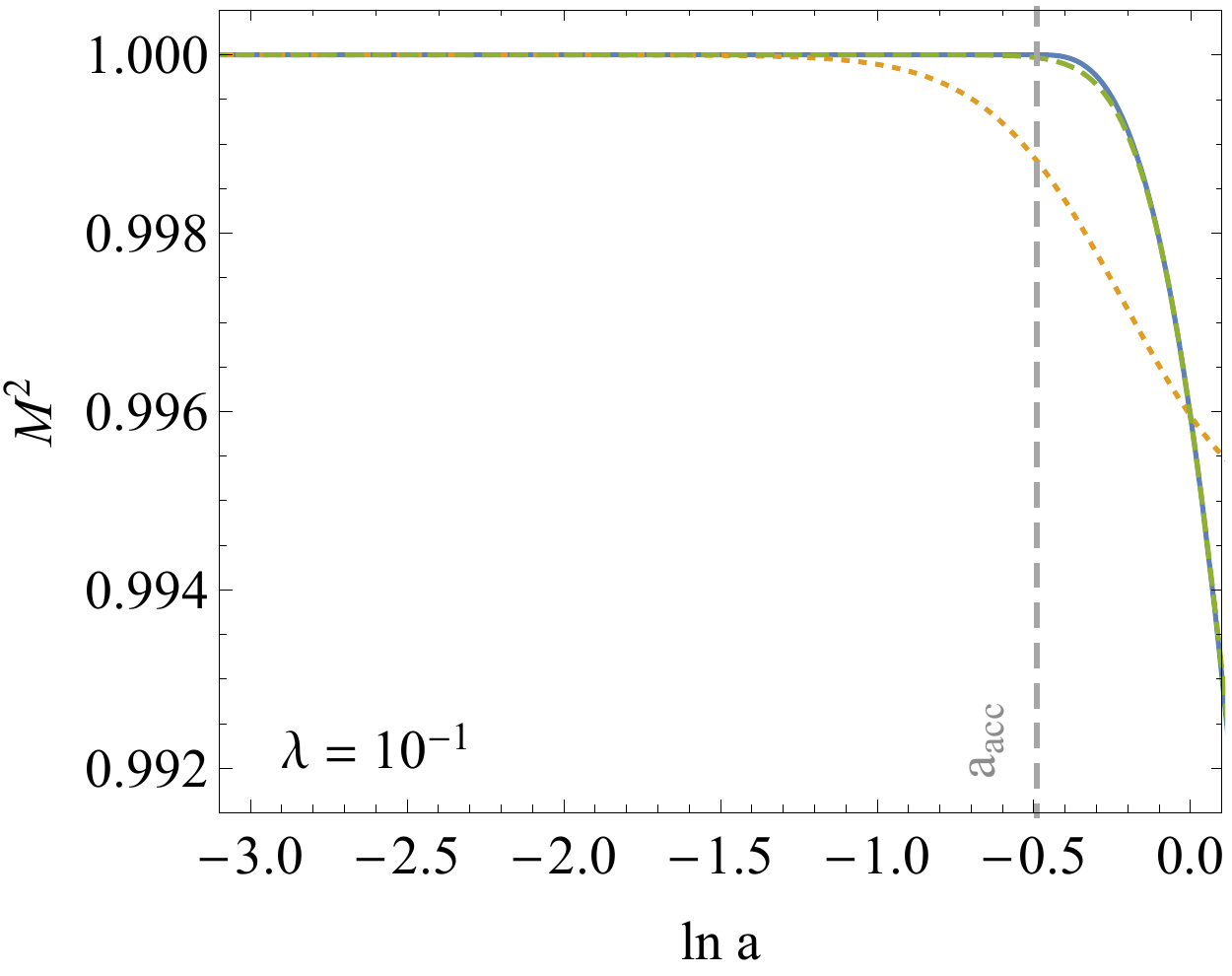}
 }
\caption{Simple generic approximations of the running Planck mass from combining different limits of the evolution rate in a broken power law, illustrated for $f(R)$ gravity (\emph{left panel}) and the extended minimal self-acceleration model shown in Fig.~\ref{fig:minselfacc} (\emph{right panel}).
Solid curves depict the exact solutions whereas dotted and dashed curves represent the combination of the future evolution of $M^2$ with the past limit and with its current evolution rate, respectively.
Performing the approximation that employs the current evolution rate in particular, one can evade the integration of Eq.~\eqref{eq:diffeq} to increase numerical efficiency.
}
\label{fig:M2approximations}
\end{figure}

We illustrate the different approximations in Fig.~\ref{fig:M2approximations} and find that the exponents determined from Eq.~\eqref{eq:malt} exhibit a sufficiently accurate match to the exact solutions of $M^2$ for observational applications.

Finally, we emphasise that there is an important difference in the parametrisation of $\alpha_i$ functions as approximations of their exact solutions for a given model and the approximation of $M^2$ or $\aB$ as the prediction of a given stable set of parameters.
The approximated predictions only need to be accurate within the observational constraint that can be inferred from the data.
In contrast, the approximations of the $\alpha_i$ functions are used to evaluate whether the particular set of parameters involved is stable or not.
This stability check can lead to the refusal of an entire class of viable theories simply because the error in the approximation yields a small violation of the stability criteria or a particular theoretical prior that is imposed, for instance, a small superluminal excess in the sound speed for a theory where the sound speed is strictly $\cs=1$.

\subsection{Phenomenology of the modifications} \label{sec:phenomenology}

Before concluding our analysis, we shall briefly inspect the limitations that the requirement of stability places on the possible phenomenology of dark energy and modified gravity models from the perspective of the manifestly stable basis.
For this purpose, we cast the effective modification of the Poisson equation and the gravitational slip, Eqs.~(\ref{eq:muQS}) and (\ref{eq:gammaQS}), in the quasistatic regime at $z=0$ into the stable basis.
In terms of the parametrisation introduced in Sec.~\ref{sec:parametrisation}, these take a very simple form:
\begin{eqnarray}
 \mu_{0} & \equiv & \mu_{\infty}(z=0) = \left. \frac{1}{M^2} \frac{2 (\aB - \aM)^2 + \alpha\cs^2}{\alpha\cs^2} \right|_0= \frac{1}{M_0^2} \frac{2 (1 - b)^2 \aBtoday^2 + \alpha_0(1+s)}{\alpha_0(1+s)} \,, \label{eq:mu} \\
 \gamma_0 & \equiv & \gamma_{\infty}(z=0) = \left. \frac{2\aB(\aB - \aM) + \alpha\cs^2}{2(\aB-\aM)^2 + \alpha\cs^2} \right|_0 = \frac{2(1 - b)\aBtoday^2 + \alpha_0(1+s)}{2(1-b)^2\aBtoday^2 + \alpha_0(1+s)} \label{eq:gamma} \,.
\end{eqnarray}
To analyse and illustrate the available model space, we shall further parametrise $\alpha_0=m\aBtoday^2$, which is directly motivated from consideration of Jordan-Brans-Dicke gravity, where $\alpha_0=2(2\omega_0+3)\aBtoday^2$ (Sec.~\ref{sec:JBD}).
Hence, we find
\begin{eqnarray}
 \mu_0 & = & \frac{1}{M_0^2} \frac{2 (1 - b)^2 + m (1+s)}{m (1+s)} = \frac{1}{M_0^2} \frac{2 + x}{x} \,, \label{eq:muspace} \\
 \gamma_0 & = & \frac{2(1 - b) + m (1+s)}{2(1-b)^2 + m (1+s)} = \frac{2/(1 - b) + x}{2 + x} \label{eq:gammaspace} \,,
\end{eqnarray}
where we have furthermore defined $x\equiv m(1+s)/(1-b)^2>0$.

\begin{figure}
 \centering
 \resizebox{0.496\textwidth}{!}{
 \includegraphics{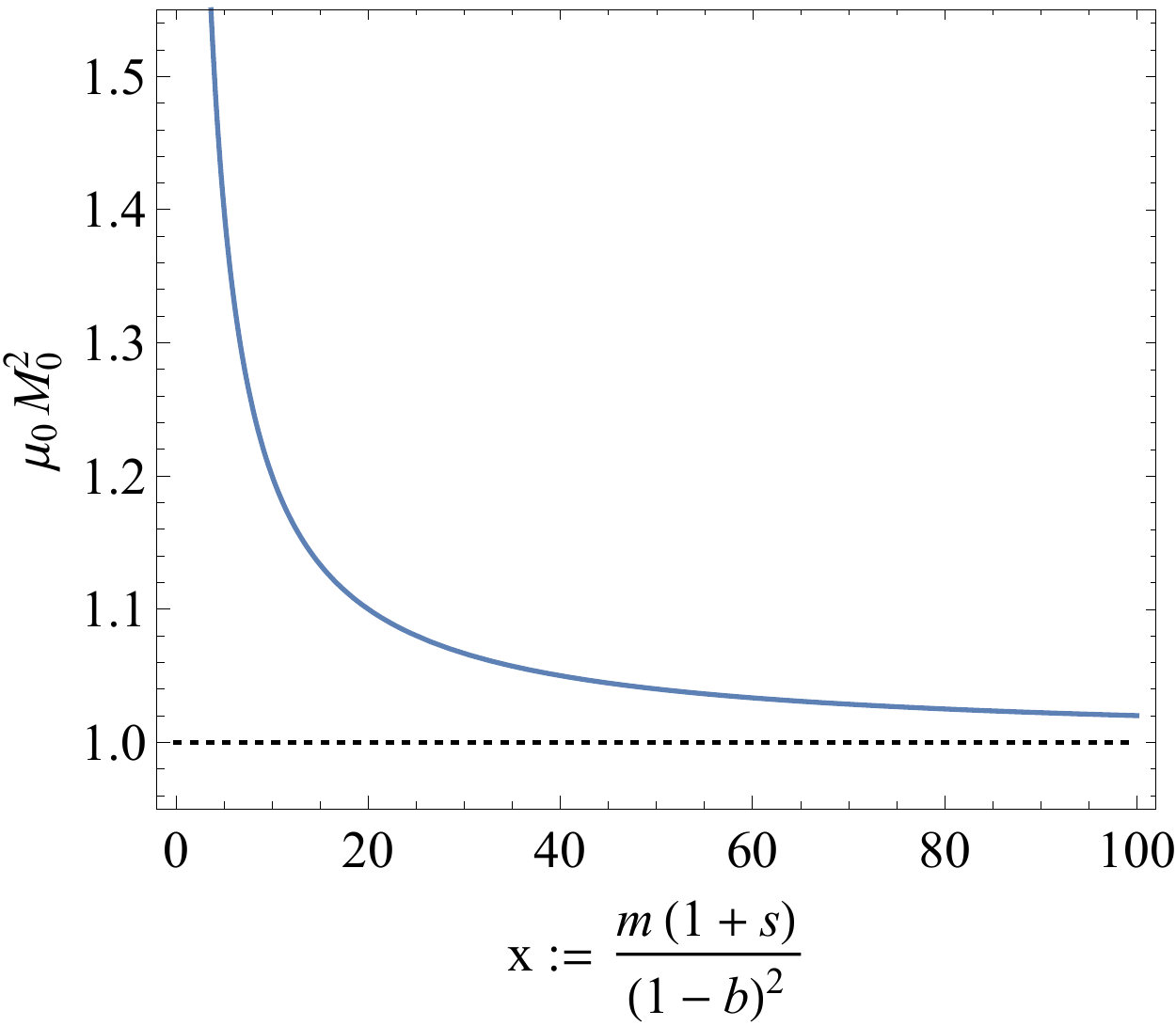}
 }
 \resizebox{0.496\textwidth}{!}{
 \includegraphics{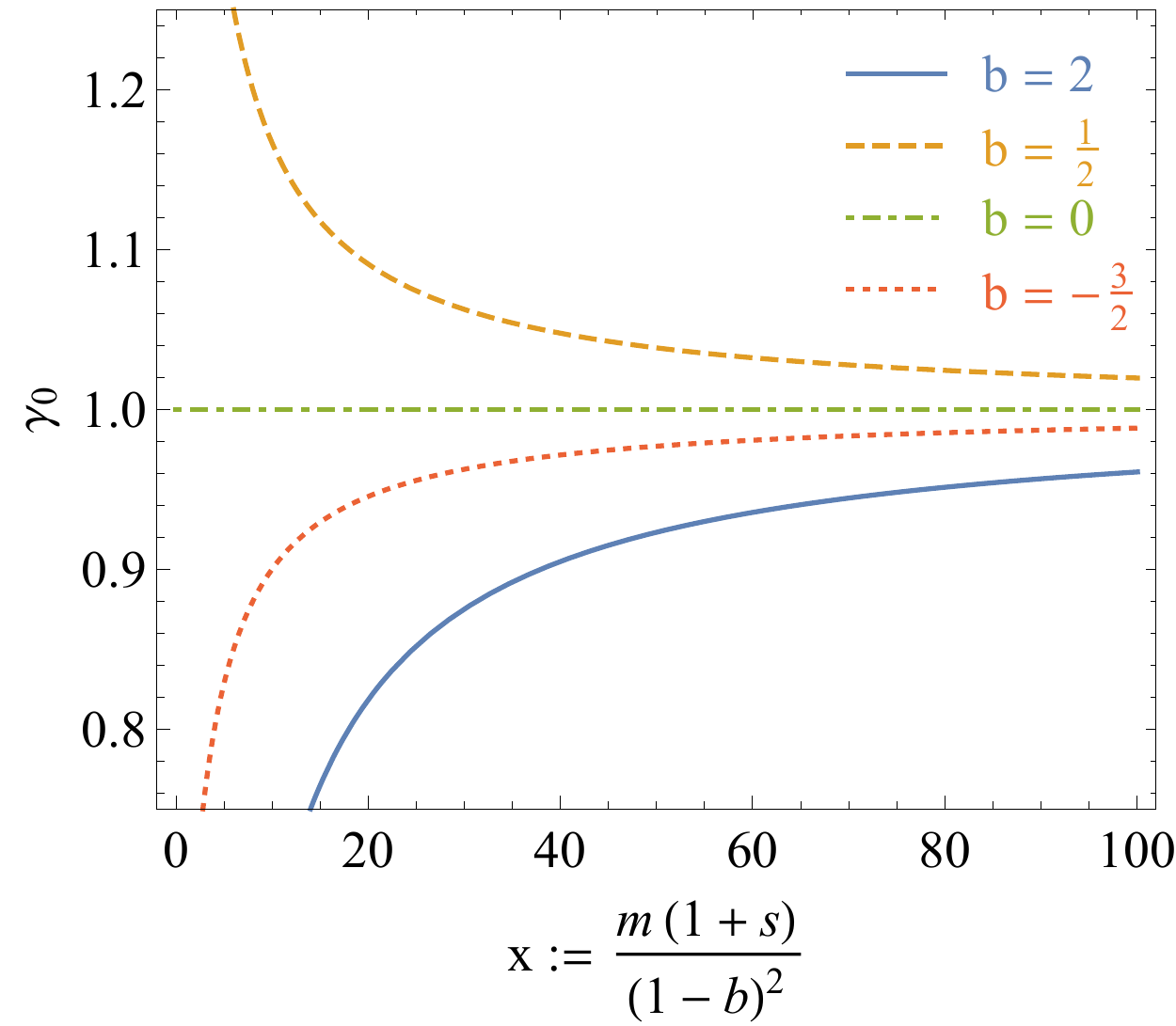}
 }
\caption{Stable model space generated by the parametrisation of the stable basis defined by Eqs.~\eqref{eq:w0wa}--\eqref{eq:alphaparam} and Eqs.~\eqref{eq:muspace}--\eqref{eq:gammaspace}.
\emph{Left panel:} The present modification of the Poisson equation $\mu_0$ is driven to $M_0^{-2}$ for large $x$. For $0<M_0^2<1$ and $M_0^2>1$ this corresponds to enhanced and weakened gravity, respectively.
\emph{Right panel:} The present gravitational slip $\gamma_0$ is larger than unity for $0<b<1$ and smaller otherwise.
The GR values $\mu_0=\gamma_0=1$ are recovered in the limit of $x\rightarrow\infty$ (or $b\rightarrow1$) and $M_0^2=1$.
The stability conditions~\eqref{eq:stability} do not restrict the values that $\mu_0$ and $\gamma_0$ can take besides excluding the anti-gravity regime $M_0^2<0$.
}
\label{fig:modelspace}
\end{figure}

Fig.~\ref{fig:modelspace} shows the range of $\mu_0$ and $\gamma_0$ as functions of $x$ for different values of $b$ and $M_0^2$.
Note that when $b=1$, we simply have $\gamma_0=1$ and $\mu_0=M_0^{-2}$.
From our definition of $x$ we cannot strictly set $b=1$ but the modifications are reproduced in the limit where $b\rightarrow1$ with $x\rightarrow\infty$ for constant $m$ and $s$.
For $0<b<1$, one finds that $\gamma_0>1$, and weak gravity $\mu_0<1$ for $M_0^2>(2+x)/x$.
Otherwise, more generally, we have $\gamma_0\leq1$ and enhanced gravity $\mu_0\geq1$.
The GR values $\gamma_0=\mu_0=1$ are reproduced here in the limit of $x\rightarrow\infty$ ($b\rightarrow1$) and $M_0^2=1$.
It is worth noting that due to the breaking of degeneracies in the model predictions by the $\cT=1$ constraint, the GR values cannot be reproduced by a modified gravity model with an evolving $M^2\neq1$~\cite{Lombriser2015c}.

Importantly, we find here that there are no general limitations on the current values that the phenomenological modification of the Poisson equation $\mu_0$ and the gravitational slip $\gamma_0$ can take from stability requirements besides the exclusion of anti-gravity with $\mu_0<0$, as illustrated in Fig.~\ref{fig:modelspace}.

Finally, one may wish to cast the parametrisation in Sec.~\ref{sec:parametrisation} into some measurable quantities that are closer to observations, for instance, in terms of $\mu_0$ and $\gamma_0$.
For this purpose, we can invert Eqs.~\eqref{eq:mu} and \eqref{eq:gamma} for $M_0^2$ and $\aBtoday$ to find
\begin{eqnarray}
 M_0^2 & = & \frac{b}{\mu_0[1 + (b-1)\gamma_0]} \,, \\
 \aBtoday & = & \pm \frac{1}{\sqrt{2}} \sqrt{\frac{\alpha_0(1-\gamma_0)(1+s)}{b-1 + (b-1)^2\gamma_0}} \,.
\end{eqnarray}
Hence, for a choice of $b\neq1$ and $\alpha_0$, the phenomenological modifications $\gamma_0$ and $\mu_0$ can be directly incorporated into the parametrisation in Sec.~\ref{sec:parametrisation}.
Note again that for $b=1$, we simply have $\gamma_0=1$ and $\mu_0=M_0^{-2}$.

\section{Conclusions} \label{sec:conclusions}

Cosmological observations test gravity on vastly different length scales than the conventional probes in the Solar System and are indispensable for reaching an understanding of the dark sector, in particular of dark energy and the late-time accelerated expansion of our Universe. A growing wealth of observational data on the cosmic background, the large-scale structure, and the propagation of gravitational waves enables to place succeedingly more stringent constraints on dark energy and modified gravity models. To most effectively exploit this gain in cosmological information and infer the broadest of implications for the vast model space available, it is important to develop unified frameworks that enable generalised and efficient predictions for the observables. The employment of the EFT of dark energy and modified gravity has been particularly successful for this purpose. But in its practical application the formalism has so far still been restrained by questions surrounding the adequate parametrisation of the free time-dependent functions inherent to the framework, which are demanded to expose a multitude of useful features.
These range from simplicity, representativity of known theories, and the theoretical soundness of the sampled model space to the efficiency in the prediction of observables required for extensive numerical applications.

In this paper, we have developed an inherently stable EFT formalism that covers the phenomenology of the cosmic background and linear fluctuations generated by Horndeski scalar-tensor theories with luminal speed of gravitational waves.
It can easily be configured to evade theoretical pathologies such as ghost or gradient instabilities in the sampled model space but also easily accommodates further theoretical priors such as the requirement of luminal or subluminal sound speed of the scalar field fluctuation.
We have found that the new EFT basis can conveniently represent a broad range of known dark energy and gravitational theories and that the time dependence of the basis functions can furthermore be parametrised in a straightforward fashion to provide a simple but general representation of the wide model space.
We also explored how approximations to the phenomenological modifications generated by the EFT basis can be approximated to efficiently predict observations and be implemented in parameter estimation analyses. We furthermore identified that the values for the current modification of the Poisson equation and the gravitational slip due to the inclusion of a new scalar degree of freedom coupled to the metric or matter fields generally remains unconstrained by stability requirements, except for the exclusion of anti-gravity.

We believe that with the introduction and parametrisation of the stable EFT basis, we are well prepared for inferring generalised and physically meaningful constraints on the space of modified gravity and dark energy models with current~\cite{Aghanim:2018eyx,Hildebrandt:2016iqg,Abbott:2017wau,Abbott2017c} and future~\cite{Ivezic2008,Laureijs2011,Amaro-Seoane:2017} cosmological data from the linear regime, but there is room for further improvement in the numerical efficiency of the theoretical predictions of observables. We leave the implementation of the inherently stable EFT in a parameter estimation analysis to future work.

\section*{Acknowledgments}

L.L.~and C.D.~were supported by a Swiss National Science Foundation (SNSF) Professorship grant (No.~170547).
J.K.~acknowledges funding through a Science and Technology Facilities Council (STFC) studentship.
A.N.T.~thanks the Royal Society for support from a Wolfson Research Merit Award and acknowledges funding by the STFC Consolidated Grant for Astronomy and Astrophysics at the University of Edinburgh.
Please contact the authors for access to research materials.

\appendix
\section{Mappings} \label{sec:mappings}

For convenience and for use to further applications, we provide in Sec.~\ref{sec:horndeski2isb} the equations mapping Horndeski scalar-tensor theories with luminal tensor sound speed to the stable EFT basis, and in Sec.~\ref{sec:isb2horndeski} we present the reconstruction of $\cT=1$ Horndeski models from a given set of stable basis functions.

\subsection{From Horndeski gravity to inherently stable effective field theory} \label{sec:horndeski2isb}

Given a covariant Horndeski scalar-tensor theory with functions $G_i$ in Eq.~(\ref{eq:horndeski}) and luminal speed of gravitational waves, the stable EFT basis of Sec.~\ref{sec:EFTISB} is found by evaluating the relations
\begin{eqnarray}
 M^2 & = & 2 G_4 \,, \\
 H^2 M^4 \alpha & = & 4 X \left\{ \vphantom{\dot{\phi}} 3 (G_{4\phi} + G_{3X} X)^2 + 
   G_4 \left[ G_{2X} + 2 \left( G_{3\phi} + \{ G_{2XX} + G_{3\phi X} \} X \right) \right] \right. \nonumber\\
   & & \left. - 6 H G_4 ( G_{3X} + G_{3XX} X ) \dot{\phi} \right\} \,. \\
   c_s^2 & = & - \left\{ G_4 \left[  \vphantom{\dot{\phi}} 4 G_4 H H' + \rhom + 4 X (G_{4\phi\phi} + G_{3\phi X} X) - 2 H ( G_{4\phi} - G_{3X} X ) \dot{\phi} \right.\right. \nonumber\\
   & & \left.\left. + 2 \left( G_{4\phi} + X \{3 G_{3X} + 2 G_{3XX} X\} \right) \ddot{\phi} \right] + 2 X (G_{3X} X - 3G_{4\phi})(G_{4\phi} + G_{3X}X)  \vphantom{\dot{\phi}} \right\} \nonumber\\
   & & \times \left\{ 2 X \right\}^{-1} \left\{ \vphantom{\dot{\phi}} 3 (G_{4\phi} + G_{3X} X)^2 + G_4 \left[ G_{2X} + 2 \left( G_{3\phi} + \{G_{2XX} + G_{3\phi X}\}  X \right) \right] \right. \nonumber\\
   & & \left.  - 6 G_4 H (G_{3X} + G_{3XX} X) \dot{\phi} \right\}^{-1} \,.
\end{eqnarray}
Furthermore, $\aM=(\ln M^2)'$, and in the case of a genuine modification of gravity with $\aM\neq0$, we can find $\dot{\phi} = H\aM G_4/G_{4\phi}$.
Also note that $H M^2\aB = \dot{\phi}(X G_{3X} + G_{4\phi})$.
Hence, similarly one may use $\dot{\phi}=(2H\aB/G_{3X})^{1/3}$ for models with $\aB\neq0$ and $M^2=1$.

The two Friedmann equations are
\begin{eqnarray}
 3 M^2 H^2  & = & \rhom -G_2 + 2 (G_{2X} + G_{3\phi}) X - 6 H (G_{4\phi} + G_{3X} X) \dot{\phi} \,, \\
 M^2 (H^2)' + 3 M^2 H^2 & = & -G_2 - 2 \left[ (G_{4\phi} + G_{3X} X) \ddot{\phi} + 2 H G_{4\phi}\dot{\phi} + X (G_{3\phi} + 2 G_{4\phi\phi}) \right] \,,
\end{eqnarray}
where we can conveniently define $3H^2=\rhom + \rhoeff$ such that
\begin{equation}
 \frac{(H^2)'}{H^2} = -3 \Om(a) - 3 [1-\Om(a)][1+\weff(a)] \,,
\end{equation}
using the energy conservation equation $\rhoeff'=-3[1+\weff(a)]\rhoeff$.
Here $\rhoeff$ defines the energy density associated with the non-GR and dark energy terms.
In Secs.~\ref{sec:models} and \ref{sec:parametrising}, we simply referred to this as $\rhoDE$, although these terms may generally not share the properties of a dark energy.

\subsection{From inherently stable effective field theory to Horndeski gravity} \label{sec:isb2horndeski}

Finally, we present the reconstruction of the covariant $\cT=1$ Horndeski action from the stable EFT basis following Ref.~\cite{Kennedy2017}.
The reconstructed Horndeski functions are
\begin{eqnarray}
G_{2}(\phi, X) & = & -U(\phi) + Z(\phi)X+4a_{2}(\phi)X^{2} + \Delta G_{2} \,, \label{eq:G2}
\label{eq:G2recon} \\
G_{3}(\phi,X) & = & \: b_{0}(\phi)-2b_{1}(\phi)X+\Delta G_{3} \,,
\label{eq:G3recon} \\
G_{4}(\phi, X) & = & \: \frac{1}{2}F(\phi) \,,
\label{eq:G4recon} \\
G_{5}(\phi, X) & = & 0 \,, \label{eq:G5}
\label{eq:G5recon}
\end{eqnarray}
where any contribution to $b_{0}(\phi)$ can be absorbed into $Z(\phi)$ through an integration by parts, and recall that the Planck mass has been set to unity.
As this Horndeski action is reconstructed from the background and linear perturbations only, there remains a great deal of freedom in the nonlinear regime, where different gravitational theories that are degenerate at the linear level may depart from each other.
This freedom is characterized by the $\Delta G_{i}$ terms which are corrections that can be added onto the reconstructed $G_i$ functions to move between different Horndeski theories with luminal tensor sound speed without changing linear cosmological predictions.
These correction terms may be specified as   
\begin{equation}
 \Delta G_{2,3} = \sum_{n>2} \xi^{{\scriptscriptstyle(2,3)}}_{n}(\phi)\left(1-2X\right)^{n} \,, \\
\end{equation}
where $\xi^{{\scriptscriptstyle(i)}}_{n}(\phi)$ are arbitrary functions of the scalar field $\phi$.
Note that the limits in the summation are set such that, in the unitary gauge, any contribution from these terms will enter at least as third-order perturbations in the EFT action~\eqref{eq:eftaction} and hence only take effect beyond the linear cosmological perturbations.

In terms of the stable EFT basis, assuming that $\aB$ has been determined from Eq.~\eqref{eq:secondorderODE}, the reconstructed action defined by Eqs.~\eqref{eq:G2}--\eqref{eq:G5} can be expressed as 
\begin{eqnarray}
    U(\phi) & = & HM^{2}\left[\left(-\alpha+24+6\aB(3+\aB)+3\aM(4+\aM)+3\aM'\right)H+\left(6+3\aM \right)H' \right] \nonumber\\
    & & -\frac{1}{8}\left(\rhom+\tilde{\beta}(a)\right)\, , \\
    Z(\phi) & = &-\frac{HM^{2}}{2}\left[\left(\alpha-6\aB(1+\aB)+3\aM^{2}+3\aM' \right)H + 3H'\left(2+\aM \right) \right] \nonumber\\
    & & -\frac{1}{2}\left(3\rhom - \tilde{\beta}(a) \right) \,, \\
    a_{2}(\phi) & = & \frac{HM^{2}}{8}\left[ \left(\alpha-6\aB(-1+\aB)+\left(-4+\aM \right)\aM+\aM'\right)H+\left( 2+\aM \right)H'     \right] \nonumber\\
    & & +\frac{1}{8}\left(\rhom+\tilde{\beta}(a)\right) \,, \\
    b_{1}(\phi) & =& \frac{1}{2}HM^{2}(\aM-2\aB) \,, \\
    F(\phi) & = & M^{2} \,,
\end{eqnarray}
where we have defined
\begin{equation}
    \tilde{\beta}(a)\equiv HM^{2}\aM \left[\left(\aM-2\aB\right)H\right]'
\end{equation}
for compactness.
Note that $\tilde{\beta}=0$ if $b_1=0$ or $G_3=0$.

\bibliographystyle{JHEP}
\bibliography{stableparam}

\end{document}